\documentclass[letterpaper,twocolumn,10pt]{article}
\usepackage{usenix-2020-09}
\usepackage{tikz}
\usepackage{amsmath}

\usepackage{soul}
\usepackage{marginnote}
\usepackage[utf8]{inputenc}
\usepackage[normalem]{ulem}
\usepackage{amsfonts} 
\usepackage{fontawesome}
\usepackage{amssymb}
\usepackage[ruled,lined,linesnumbered]{algorithm2e} 
\usepackage{soul}
\usepackage{algorithmic}
\usepackage{setspace}
\usepackage{graphicx}
\usepackage{multirow}
\usepackage{booktabs}
\usepackage{bbm}
\usepackage{tikz}
\usetikzlibrary{positioning}

\usetikzlibrary{decorations.pathreplacing, automata}
\usetikzlibrary{arrows}
\usepackage{amsmath}
\usepackage{pifont}

\usetikzlibrary{shapes}
\makeatletter
\tikzset{circle split part fill/.style args={#1,#2}{%
 alias=tmp@name, 
  postaction={%
    insert path={
     \pgfextra{%
     \pgfpointdiff{\pgfpointanchor{\pgf@node@name}{center}}%
                  {\pgfpointanchor{\pgf@node@name}{east}}%
     \pgfmathsetmacro\insiderad{\pgf@x}
      \fill[#1] (\pgf@node@name.base) ([xshift=-\pgflinewidth]\pgf@node@name.east) arc
                          (0:180:\insiderad-\pgflinewidth)--cycle;
      \fill[#2] (\pgf@node@name.base) ([xshift=\pgflinewidth]\pgf@node@name.west)  arc
                           (180:360:\insiderad-\pgflinewidth)--cycle;            
         }}}}}
\usetikzlibrary{spy}
\usetikzlibrary{arrows}
\usepackage{amsmath}
\usepackage{pifont}

\usepackage{siunitx}
\usepackage{amsthm}
\usepackage{thmtools,thm-restate}

\usepackage{mdframed}

\usepackage{subfigure}

\usepackage{pgfplots}

\usepackage[inline]{enumitem}
\setlist[enumerate]{noitemsep, topsep=2pt}
\setlist[itemize]{noitemsep, topsep=2pt}
\setlist[description]{noitemsep, topsep=2pt, font=\normalfont\space}

\usepackage{pgfplots}
\pgfplotsset{compat=1.15} 
\pgfplotsset{%
    axis line origin/.style args={#1,#2}{
        x filter/.append code={ 
            \ifx\pgfmathresult\empty\else\pgfmathparse{\pgfmathresult-#1}\fi
        },
        y filter/.append code={
            \ifx\pgfmathresult\empty\else\pgfmathparse{\pgfmathresult-#2}\fi
        },
        xticklabel=\pgfmathparse{\tick+#1}\pgfmathprintnumber{\pgfmathresult},
        yticklabel=\pgfmathparse{\tick+#2}\pgfmathprintnumber{\pgfmathresult}
    }
}
\usepackage{xspace}
\usepackage{numprint}

\usepackage{balance}

\npthousandsep{,}

\usepackage{url}
\usepackage{color}
\usepackage{etoolbox}
\makeatletter
\patchcmd{\@makecaption}{\scshape}{}{}{}
\makeatother
\newif\ifshowcomment
\showcommentfalse

\newtheorem{definition}{Definition}

\newcommand{\Proof}{\mathit{moc}\xspace}

\newcommand{\Verify}{\nu\xspace}
\newcommand{\accuser}{\mathcal{A}\xspace}
\newcommand{\suspect}{\mathcal{S}\xspace}
\newcommand{\judge}{\mathcal{J}\xspace}

\newcommand{\xx}{\mathbf{x}\xspace}
\newcommand{\yy}{\mathbf{y}\xspace}

\newcommand{\bu}[1]{{\bf \underline{#1}}}
\newcommand{\bo}[1]{{\bf #1}}

\newcommand{\fig}{\textrm{Figure}\xspace}

\newcommand{\Paragraph}[1]{~\vspace*{-0.8\baselineskip}\\{\bf #1}}

\newcommand{\revision}[1]{{{{#1}}}}

\newcommand{\old}[1]{{}}
\newcommand{\new}[1]{{#1}}

\newcommand{\margintext}[1]{\footnotesize{\textbf{\textcolor{brown}{}}}}

\begin{document}

\date{}

\title{\Large \bf False Claims against Model Ownership Resolution}
\author{
        Jian Liu\thanks{Co-first authors; Jian Liu is the corresponding author.}$^*$\\
        Zhejiang University\\
        jian.liu@zju.edu.cn
        \and
        {\rm Rui Zhang$^*$}\\
        Zhejiang University \\
        zhangrui98@zju.edu.cn
        \and
        {\rm Sebastian Szyller}\\
        Intel Labs \&Aalto University \\
        contact@sebszyller.com 
        \and
        {\rm Kui Ren}\\
        Zhejiang University \\
        kuiren@zju.edu.cn
        \and
        {\rm N. Asokan}\\
        University of Waterloo \& Aalto University \\
        asokan@acm.org
    } 
\maketitle

\begin{abstract}
Deep neural network (DNN) models are valuable intellectual property of model owners, constituting a competitive advantage.
Therefore, it is crucial to develop techniques to protect against model theft.
Model ownership resolution (MOR) is a class of techniques that can deter model theft. A MOR scheme enables an \emph{accuser} to assert an ownership claim for a \emph{suspect model} by presenting evidence, such as a watermark or fingerprint, to show that the suspect model was stolen or derived from a source model owned by the accuser.
Most of the existing MOR schemes prioritize robustness against malicious suspects,  ensuring that the accuser will win if the suspect model is indeed a stolen model.

In this paper, we show that common MOR schemes in the literature are vulnerable to a different, equally important but insufficiently explored, robustness concern: a \emph{malicious accuser}. We show how malicious accusers
can successfully make \emph{false claims} against \emph{independent} suspect models that were not stolen.
Our core idea is that a malicious accuser can deviate (without detection) from the specified MOR process by finding (transferable) adversarial examples that successfully serve as evidence against  independent suspect models.
To this end, we first 
generalize the  procedures of common MOR schemes 
and show that, under this generalization, defending against false claims is as challenging as preventing (transferable) adversarial examples.
{Via systematic empirical evaluation}, we show that our false claim attacks always succeed in \old{all prominent MOR schemes with realistic configurations}\new{MOR schemes that follow our generalization}\marginpar{\margintext{Review-C-2}}, 
including in a real-world model: Amazon's Rekognition API.

\end{abstract}

\section{Introduction}

Deep Neural Networks (DNNs) have been used extensively in many real-world applications such as facial recognition \cite{parkhi2015deep, wang2018cosface, liu2017sphereface}, medical image classification \cite{zhang2019medical} and autonomous driving \cite{luo2017traffic}. 
However, training DNNs is expensive due to the high costs of preparing training data and fine-tuning models.
Therefore, DNNs confer a competitive advantage
to model owners who would like to prevent theft and unauthorized redistribution of their {\em source models}. 
A thief may deploy a {\em stolen model} for profit. 
The stolen model can be an exact copy of the source model (with possible subsequent refinement) or a surrogate model \emph{extracted} by querying the source model's inference interface.


\Paragraph{Model ownership resolution (MOR).}
Preventing model stealing is difficult \cite{krishna2020thieves, jagielski2020high, carlini2020cryptanalytic, atli2020extraction}, 
but mechanisms for detecting a stolen model and resolving ownership serve as powerful deterrents.
A {\em model ownership resolution} (MOR) scheme enables an {\em accuser} ($\accuser$) to present evidence, such as a watermark or fingerprint, to a {\em judge} ($\judge$) and claim that a model held by a {\em suspect} ($\suspect$) is a stolen model.

DNN watermarking~\cite{Adi, zhang, li2019piracy} is one type of MOR;
it embeds a watermark into a DNN during training.
A watermark typically consists of a set of samples with incorrectly assigned labels, known as the {\em trigger set}. 
The model owner uses the {trigger set}, along with the training set, to train the source model. 
The watermarked source model, and any model derived from it, will perform differently on the trigger set compared to other \emph{independent}\footnote{Henceforth, we use ``independent models'' to represent models that are independently trained for the same task as $\accuser$'s source model.} models.
Therefore, the trigger set can help to resolve the ownership.

DNN fingerprinting is another type of MOR,
which extracts a unique identifying code (fingerprint) from an already trained model.
Unlike watermarking, which embeds a watermark into a DNN during training, fingerprinting does not require any changes to the training phase and hence does not sacrifice model accuracy. Similar to watermarking, verification of a fingerprint involves querying a suspect model using a trigger set that corresponds to the fingerprint.

\Paragraph{False claims against MOR.}
Most of the existing MOR schemes prioritize the robustness against a malicious suspect: they try to make sure that the accuser will win the case if the suspect model is indeed a stolen model.
In this paper, we focus on a different under-explored robustness problem: that of a \emph{malicious accuser}. 
Specifically, \ul{we investigate whether a malicious accuser can falsely claim ownership of an indepdent suspect model that is {\em not} a stolen model}.

We posit that the robustness of a MOR scheme against malicious accusers is just as crucial 
as its robustness against malicious suspects. 
Indeed, if a malicious accuser can falsely claim  ownership of independent models, the corresponding MOR scheme will not be useful in settings that need to resolve \old{model ownership legally}\new{legal model ownership}.\marginpar{\margintext{Review-C-3}} 

The judge's MOR decision is based on the inference results of a suspect model on the trigger set.
Our intuition is that \ul{a malicious accuser can construct a trigger set in a way to raise false positives for all independent models}. 
This can be achieved via (the transferability of) adversarial examples.
To demonstrate the ubiquity of this attack, we first generalize the procedures for common MOR schemes, and then show that any MOR scheme following our generalization is susceptible to this attack.
We survey 16 MOR schemes and \old{\emph{show that all} of them follow our generalization.}\new{show that our generalization can effectively capture these surveyed schemes.}
\footnote{Parameter-encoding watermarking schemes do not conform our generalization. However they are known to be {\em not} robust~\cite{wang2019attacks}. (cf. Section~\ref{sec:related})}
\marginpar{\margintext{Review-C-2}}
%
We {empirically evaluate several well-known MOR schemes}~\cite{Adi, Jia, DI, DAWN, Lukas} to show that all of them are vulnerable to our false claim attacks. 
Specifically, we attempt to falsely claim models trained from CIFAR-10 and ImageNet, as well as the model behind Amazon Rekognition API~\cite{Amazon}.
Our evaluation shows that, in realistic configurations, our false claims can succeed against all \new{evaluated}\marginpar{\margintext{Review-C-2}} MOR schemes for all models.



We summarize our contributions as follows:
\begin{itemize}
    \item We present a 
    \new{generalization of MOR procedures},
    and an adversary model for malicious accusers. (Section~\ref{sec:mor}) We show that all secure MOR schemes 
    follow our generalization\old{.}\new{;}  (Section~\ref{sec:survey})
    \marginpar{\margintext{Review-E-1}}
    \item 
    We show that 
    \new{false claims against these schemes can always succeed}, unless (transferable) adversarial examples can be prevented\old{.}\new{;} (Section~\ref{sec:attack})
    \item We 
    \new{empirically evaluate these schemes including on a real-world model held by Amazon}, to
    show that 
    our false claim attacks do succeed\footnote{Source code at \url{https://github.com/ssg-research/Falseclaims}}\old{.}\new{;}  (Section~\ref{sec:eval})
    \item 
    We 
    \new{provide some guidance on augmenting MOR schemes to withstand our attacks}. (Section~\ref{sec:countermeasures})
\end{itemize}
\begin{table}[tb]
\new{\caption{Summary of frequent notations.}}\marginnote{\margintext{Review-E-1}}
\small
\centering
\begin{spacing}{1.30}
\begin{tabular}{p{1.7cm} p{5.8cm}}
\toprule[1pt]
\textbf{Notation} & \textbf{Description} \\ \hline
\toprule[0.8pt]
$\accuser$ & Accuser \\ \hline
$\suspect$ & Suspect \\ \hline
$\judge$ & Judge \\ \hline
$F_{\accuser}$     & source model \\ \hline
$F_{\suspect}$     & suspect model \\ \hline
$\Proof_\accuser$ & a model ownership claim submitted by $\accuser$ \\ \hline
$f()$     & ground-truth function \\ \hline
$\Verify()$ & a verification function \\ \hline
$cm$ & a cryptographic commitment \\ \hline
$x$   & a sample \\ \hline
$\xx$ & a set of samples \\ \hline
$y$   & a label \\ \hline
$\yy$ & a set of labels \\ \hline
$T$ & a decision threshold \\ \hline
MOR & model ownership resolution \\ \hline
$\mathit{MORacc}$ & MOR accuracy \\  \hline
\new{$\epsilon$} & \new{pre-pixel perturbation bounds\marginnote{\margintext{Review-C-3}}} \\
\bottomrule
\end{tabular}
\end{spacing}
\label{notationtable}
\vspace{-3mm}
\end{table}
\section{Preliminaries}

In this section, we provide some fundamental concepts to facilitate the  understanding of this paper.

\subsection{Deep neural networks (DNN)}

A {\em deep neural network} (DNN) model is a mathematical function $F()$ that assigns a label $y$ to a sample $x$:
$$y\leftarrow F(x).$$
A DNN consists of a series of layers, with each layer employing a linear function followed by an activation function.
In particular, the softmax activation function is commonly employed in the output layer to convert likelihood scores into class probabilities.

To train a DNN, it is necessary to specify a differentiable loss function $L$ that serves as an objective during the optimization process. 
The cross-entropy loss is a commonly used loss function in DNN training, which quantifies the discrepancy between the predicted output of the model and the ground truth label.

A black-box deployment of a DNN exposes only the API of the model: 
given an input, the DNN API returns a class or class probabilities.


\subsection{Adversarial examples}

{\em Adversarial examples} were first reported in 2013 as slightly perturbed images that nudge DNNs into making incorrect predictions.
Such adversarial examples look almost the same as  original images and were seen as a systematic vulnerability in DNNs~\cite{FGSM}. 
Since then, they have been explored extensively in both attacks~\cite{dong2018boosting, su2019one} and defences~\cite{papernot2016distillation,buckman2018thermometer,Bhagoji2018Enhancing,zheng2016improving,wang2017learning,luo2016foveationbased}. 

An adversarial example is generated by solving:
$$r=\mathop{\arg\min}_{r}L(F(x+r), y')+\alpha||r||,$$
where ${y}'$ is different from the real label ${y}$ for ${x}$.
Then, the noise $r$ can fool the DNN into predicting a wrong label (by minimizing the loss function) with imperceptible perturbations.

It is well-known that adversarial examples are {\em transferable}:
the adversarial examples generated for one model could mislead another model~\cite{papernot2016transferability, papernot2017practical, Delving}. 
Such a property can be leveraged to generate adversarial examples for black-box models.


\subsection{Model extraction}

\emph{Model extraction} is a kind of {black-box} attack aiming to obtain an \emph{extracted model} that is \emph{functionally equivalent} to the victim model.
Black-box refers to an attacker who gleans information from the victim model solely by interacting with its prediction API: they choose queries (inputs) and obtain the corresponding labels.
Using that information, they train their own model.

It is commonly assumed that the attacker does not know the exact architecture or the training data used to train the victim model.
However, they can choose an architecture suitable for the task and appropriate data to execute the attack.

Model extraction is effective in various tasks: image classification~\cite{tramer2016stealing,papernot2017practical,juuti2019prada,orekondy2018knockoff,correia2018copycat,jagielski2020highfidelity}, image translation~\cite{szyller2021ganstealing}, NLP~\cite{krishna2020thieves,wallace2020translation}, and others~\cite{he2021stealinglinks,takemura2020rnns}.

\section{MOR: Generalization}
\label{sec:mor}

In this section, we aim to generalize the procedures for common MOR schemes and define the capabilities/goals of a malicious accuser. Table~\ref{notationtable} summarizes our notations.

\subsection{MOR procedures}
\label{sec:mor_procedures}

A MOR scheme typically consists of two procedures: 
\begin{itemize}
    \item {\em claim generation} allows the owner (who can later act as an accuser $\accuser$) of a model $F_{\accuser}$  to generate a model ownership claim $\Proof_\accuser$ (watermark/fingerprint) for $F_{\accuser}$;
    \item {\em claim verification} allows a judge $\judge$ to use $\Proof_\accuser$ to determine whether a model $F_{\suspect}$ held by a suspect $\suspect$ is a stolen model that is derived from $F_{\accuser}$.
\end{itemize}
Next, we formally define these two procedures, attempting to cover most MOR schemes.

\Paragraph{Claim generation.}
Given $F_{\accuser}$, $\accuser$ creates $\Proof_\accuser$, which is formally defined as follows:
\begin{definition}
\label{def:moc}
    A model ownership claim for a model $F_{\accuser}$ is defined as $\Proof_{\accuser} = (\xx, \yy, aux, cm_\accuser)$, where $\xx$ is a set of data samples, $\yy$ is the corresponding set of labels, $aux$ is auxiliary information, and $cm_\accuser$ is a cryptographic commitment of $(\accuser, F_{\accuser}, \xx, \yy, aux)$ which receives a secure timestamp that can be verified by $\judge$, for example by posting the commitment on a timestamped public bulletin board. The \emph{trigger set} is
    $(\xx, \yy)$.
\end{definition}

We remark that most MOR schemes (except DAWN~\cite{DAWN}) do not explicitly mention that the commitment is timestamped, and some do not even mention commitments at all.
We add timestamped commitments to our definition for two reasons:
\begin{itemize}
    \item Without them, false claims are easier (cf. Section~\ref{sec:discuss}). 
    We consider the most difficult setting for our attack.
    \item Integrating them into existing MOR schemes is straightforward.
\end{itemize}

For a verification function $\Verify$, the claim should satisfy:
\old{$\forall (x_i, y_i)\in(\xx,\yy)$,
$\Verify(F_{\accuser}(x_i), y_i)=1$ for $F_{\accuser}$ 
and $\Verify(F(x_i),y_i)=0$ for other $F\neq F_{\accuser}$.}\new{for $F_{\accuser}$,$\forall (x_i, y_i)\in(\xx,\yy)$,
$\Verify(F_{\accuser}(x_i), y_i)=1$ and for other $F\neq F_{\accuser}$, $\Verify(F(x_i),y_i)=0$.}
\marginpar{\margintext{Review-E-1}}
For example, $\accuser$ could achieve this in  following ways:
\begin{itemize}
    \item Sample $(x_i, y_i)$ s.t. $\Verify(F(x_i),y_i)=0$ for all $F$s, and fine-tune $F_{\accuser}$ on $(x_i, y_i)$ s.t. $\Verify(F_\accuser(x_i),y_i)=1$. 
    DNN watermarking~\cite{Adi, zhang, li2019piracy, guo2018watermarking} is in this category.
    \item Exploit $F_{\accuser}$ to generate $(x_i, y_i)$ (e.g., as an adversarial example) s.t. $\Verify(F_{\accuser}(x_i), y_i)=1$ for $F_{\accuser}$, and $\Verify(F(x_i),y_i)=0$ for other $F\neq F_{\accuser}$. 
    DNN fingerprinting~\cite{Lukas, Zhao, IPGuard, Metav} falls into this category.
\end{itemize}

\Paragraph{Claim verification.}
Suppose $\accuser$ uses $\Proof_\accuser$ to claim that a model $F_{\suspect}$ (held by a suspect $\suspect$) was derived from $F_{\accuser}$.
The judge $\judge$ checks the followings:
\begin{itemize}
    \item $\Verify(F_{\accuser}(x_i),y_i)=1$, $\forall (x_i, y_i)\in(\xx,\yy)$;
    \item $\Verify(F_{\suspect}(x_i),y_i)=1$, $\forall (x_i, y_i)\in(\xx,\yy)$;\footnote{Notice that $\judge$ only needs API access to $F_{\suspect}$;  $\suspect$ need not be aware of the verification being done.}
    \item $cm_{\accuser}$ is a valid commitment of $(\accuser, F_{\accuser}, \xx, \yy)$;
    \item $cm_{\accuser}$ was timestamped before $cm_{\suspect}$ (if $cm_{\suspect}$ exists and was timestamped).
\end{itemize}


$\judge$ accepts $\accuser$'s claim if and only if all checks pass.
Recall that \old{$\Verify(F_{\accuser}(x_i), y_i)=1$ for $F_{\accuser}$, and $\Verify(F(x_i),y_i)=0$ for other $F\neq F_{\accuser}$.}\new{for $F_{\accuser}$, $\Verify(F_{\accuser}(x_i), y_i)=1$, and for other $F\neq F_{\accuser}$, $\Verify(F(x_i),y_i)=0$.}
\marginpar{\margintext{Review-E-1}}

Therefore, $\Verify(F_{\suspect}(x_i),y_i)=1$ $\forall (x_i, y_i)\in(\xx,\yy)$ only if $F_{\suspect}$ was derived from  $F_{\accuser}$.

In fact, the first and second checks do not need \new{to}\marginpar{\margintext{Review-C-3}} hold for all $(x_i, y_i)\in(\xx,\yy)$.
Instead, we define the MOR accuracy based on the watermark accuracy proposed in \cite{sok}:
\begin{equation}\label{MORacc}
    \mathit{MORacc}_F=\frac{1}{|\xx|}\sum_{i=1...|\xx|}\mathbb{I}(\Verify(F(x_i),y_i)).
\end{equation}
We say the check holds if $\mathit{MORacc}_F$ is higher than a {\em decision threshold} $T$
(cf. Section~\ref{sec:thresholds} for further discussions on $T$).

The main characteristic of this generalization is that $\Proof_\accuser$ includes a trigger set and its verification requires running model inference on the trigger set.
All MOR schemes we discuss (Section~\ref{sec:survey}) follow our generalization.



\subsection{False claims}
\label{sec:false_claims}

A robust MOR scheme should satisfy the following:
\begin{itemize}
    \item robustness against a malicious $\suspect$, i.e., $\judge$  accepts the claim if $F_{\suspect}$ is indeed a stolen model; 
    \item robustness against a malicious $\accuser$, i.e.,
    $\judge$  rejects the claim if $F_{\suspect}$ is an independent model.
\end{itemize}
Most existing MOR schemes prioritize robustness against a malicious $\suspect$. 
We focus on studying robustness against a malicious $\accuser$, who aims to {\bf falsely claim the ownership of an independent model} $F_{\suspect}$.

We assume that \ul{$\accuser$ has access to a dataset with the same distribution as the training dataset of $F_{\suspect}$}.
This is a standard assumption in MOR literature~\cite{sok, survey1, survey2}:
it implies that a false claim will succeed against $F_\suspect$ trained from a dataset with the same distribution as the training dataset of $F_\accuser$.

The timestamped commitment requirement from Definition~\ref{def:moc} implies that $\accuser$ has to create $\Proof_\accuser$ even before  $F_\suspect$  comes into existence. 
Therefore, $\accuser$ cannot know anything about $F_\suspect$ including its parameters or hyperparameters ahead of time.

\Paragraph{False claim vs. watermark forging.}
Watermark forging~\cite{zhang, xu2019novel, guo2018watermarking} allows an attacker to forge a watermark on a given model,
creating an ambiguity for $\judge$ to determine which party has watermarked the given model.
Li et al~\cite{survey2} describe three types of watermarking forging:
\begin{itemize}
    \item Recovering watermarks and claiming ownership~\cite{xu2019novel};
    \item Adding a new watermark \cite{fan2019rethinking, ASAC};
    \item Extracting a fake watermark from the model that  acts like a real one~\cite{guo2018watermarking}.
\end{itemize}
Our proposed MOR generalization, which includes a timestamped commitment, renders such attacks ineffective. 
This is because, in case $\accuser$'s and $\suspect$'s claims are both valid, $\judge$ can easily resolve  ownership based on the timestamps. 
It is impossible for the attacker to possess an earlier timestamp since watermark forging requires knowing $F_{\suspect}$, which is  available  only after the owner commits the model with a timestamp.

Therefore, the false claims considered in this paper is stronger than watermark forging.
We aim to allow $\accuser$ to claim ownership of $F_\suspect$ even if $F_\suspect$ is timestamped. 
To this end, $\accuser$  has to generate a valid $\Proof_\suspect$ for $F_\suspect$ before $F_\suspect$ \old{being}\new{is} trained.\marginpar{\margintext{Review-C-3}}
That is why we assume $\accuser$ has neither white-box nor black-box access to $F_{\suspect}$ and knows nothing about its hyperparameters.

\subsection{Decision thresholds}
\label{sec:thresholds}

It is important to carefully choose an appropriate $T$ that balances robustness against both a malicious $\accuser$ and a malicious $\suspect$.
For example, a high $T$ makes false claims difficult, but makes it easy for a stolen model to evade detection. 
Prior work~\cite{sok, Lukas} have suggested various ways for choosing $T$:
\begin{itemize}
    \item {\bf Independent.} 
    $\judge$ trains multiple independent models,  obtains their $\mathit{MORacc}$ on $(\xx, \yy)$, and chooses the highest one as the decision threshold $T$. 
    Since independent models tend to have low $\mathit{MORacc}$, the resulting $T$ is also low.
    A stolen model is likely to have a very high $\mathit{MORacc}$, making it difficult to evade the detection. 
    On the other hand,  false claims become easy because it is even possible for an innocent independent model to have a $\mathit{MORacc}$ that is higher than $T$.
    \item {\bf Extracted.} 
    $\judge$ derives multiple extracted models from $F_{\accuser}$, obtains their  $\mathit{MORacc}$ on $(\xx, \yy)$, and chooses the lowest one as the decision threshold $T$. 
    This time, $T$ will be so high that false claims are difficult to succeed.
    However, this also means that a stolen model can be easily manipulated to have a lower $\mathit{MORacc}$ than $T$, allowing it to evade the detection.
    \item {\bf Mixed.} 
    $\judge$ calculates the average $\mathit{MORacc}$ of multiple independent and extracted models
    This can be considered as a middle-ground approach.
\end{itemize}
The mixed threshold is the most sensible choice in realistic deployments, but the extracted threshold is the least favourable to a malicious $\accuser$. 
Notice that the decision threshold should be uniform for all claims and $\judge$ needs to determine it before receiving any claim.
Therefore, to determine a decision threshold, $\judge$ needs to act as $\accuser$: trains its own $F_\accuser$, and generates its own trigger set $(\xx, \yy)$ according the MOR scheme it adopts.
\section{MOR: Survey}
\label{sec:survey}

In this section, we survey 16 MOR schemes and  describe five well-known schemes under our generalization. 

\subsection{Taxonomy}
\label{sec:taxonomy}

Table~\ref{tab:survey} lists 16 MOR schemes,
including 11 watermarking schemes collected from four survey papers~\cite{sok, survey1, survey2, survey3},
and 5 fingerprinting schemes.
The watermarking schemes are categorized according to the taxonomy proposed in~\cite{sok}:
\begin{itemize}
    \item {\bf Model-Independent.} 
    The watermark is embedded into the model functionality and is independent of the model itself.
    For example, it could be an independent backdoor embedded into the source model by adding extra samples to the training set.
    \item {\bf Model-Dependent.}
    The watermark is embedded into the model functionality and depends on the model.
    For example, it could also be a backdoor embedded into the source model, but the backdoor is generated based on the source model.
    \item {\bf Active.}
    The watermark is embedded into the model functionality during inference, which means that it is only activated when the model is used to make predictions.
    Therefore, this approach only considers situations where attackers have black-box access to the source model.    
    \item {\bf Parameter-Encoding.} 
    The watermarking schemes in this category are known to be not robust~\cite{wang2019attacks}.
    We refer to Section~\ref{sec:related} for more details.
\end{itemize}

One common characteristic of watermarking schemes is that they require modifications to the source model or its inference process. 
Such modifications inevitably result in a loss of model accuracy. 
In contrast, model {\bf fingerprinting} extracts a unique identifier (i.e., fingerprint) from the source model as its trigger set, without any changes to the model itself. 
This allows the original model to be preserved, maintaining its accuracy and functionality.

\begin{table}[ht]
\new{\caption{MOR schemes.\label{tab:survey}}}\marginnote{\margintext{Review-E-1}}
\small
\centering
\begin{tabular}{c|c}
\toprule[1pt]
{\bf Category}            
&{\bf MOR scheme}        
\\\toprule[0.8pt]
\multirow{6}{*}{\begin{tabular}[c]{@{}c@{}}Model-independent\\watermarking\end{tabular}}   
                    &    Adi~\cite{Adi}          \\ 
                    &    Zhang~\cite{zhang}         \\ 
                    &   Li (a)~\cite{li2019piracy}         \\ 
                  & Guo~\cite{guo2018watermarking}   \\
                  & Namba~\cite{Namba}               \\
                  & Xu~\cite{xu2019novel}            \\
                  \hline 
\multirow{4}{*}{\begin{tabular}[c]{@{}c@{}}Model-dependent\\watermarking\end{tabular}}   
                    &    Frontier-Stitching~\cite{FS}          \\ 
                    &    Blackmarks~\cite{BlackMarks}          \\ 
                    &    \old{Jia}\new{EWE}~\cite{Jia}         \\ 
                    &    Li (b)~\cite{ASAC}          \\ \hline
\multirow{2}{*}{\begin{tabular}[c]{@{}c@{}}Active\\watermarking\end{tabular}}   
                    &    
\multirow{2}{*}{\begin{tabular}[c]{@{}c@{}}DAWN~\cite{DAWN}\end{tabular}}

                \\ \\\hline
\multirow{4}{*}{Fingerprinting}   
                    &    Lukas~\cite{Lukas}                   \\ 
                    &   \old{Zhao}\new{AFA}~\cite{Zhao}         \\
                    &    IPGuard~\cite{IPGuard}           \\ 
                    &    Metav~\cite{Metav}            \\
                    &    DI~\cite{DI}          \\ 
\bottomrule
\end{tabular}
\vspace{3pt}
\end{table}

\subsection{Generalization}
\label{sec:general}

All the MOR schemes surveyed in Section~\ref{sec:taxonomy} follow our generalization:
the model ownership claim ($\Proof$) in each scheme includes a trigger set and the claim verification requires running the model inference on the trigger set.
Next, for each category, 
we pick exemplary schemes and describe them under our generalization.
We refer to a  scheme by its first author's name for ease of presentation, unless it is known under a different name.

\subsubsection{Model-independent: Adi~\texorpdfstring{\cite{Adi}}{}}
\label{sec:Adi}

The $\Proof$ of Adi contains a set of out-of-distribution images $\xx$.
The label $y_i$ for each $x_i$ is randomly sampled over all classes excluding its true label.
The source model $F_\accuser$ needs to be fine-tuned on $(\xx, \yy)$.
If $F_\suspect$ shows 
a similar behaviour as $F_\accuser$ on $(\xx, \yy)$, then
 $F_\suspect$ is likely derived from $F_\accuser$.

\begin{itemize}
    \item {\bf claim generation.} $\accuser$ generates $\Proof_\accuser$ as follows:
    \begin{enumerate}
        \item samples out-of-distribution $\xx$ and assigns a wrong label $y_i$ to each $x_i \in \xx$, i.e., $f(x_i)\neq y_i$ where $f()$ is the ground-truth function;
        \item fine-tunes $F_\accuser$ on $(\xx, \yy)$;
        \item commits $(\accuser, F_{\accuser}, \xx, \yy)$ as $cm_\accuser$ and timestamps.
    \end{enumerate}
    \item {\bf claim verification.} $\judge$ checks the following:
    \begin{enumerate}
        \item 
        $\frac{1}{|\xx|}\sum\limits_{i=1...|\xx|}\mathbb{I}(F_\accuser(x_i)=y_i~\mathrm{and}~f(x_i)\neq y_i)>T$;
        \item 
        $\frac{1}{|\xx|}\sum\limits_{i=1...|\xx|}\mathbb{I}(F_\suspect(x_i)=y_i~\mathrm{and}~f(x_i)\neq y_i)>T$;
        \item $cm_{\accuser}$ is a valid commitment of $(\accuser, F_{\accuser}, \xx, \yy)$;
           \item $cm_{\accuser}$ was timestamped before $cm_{\suspect}$. 
    \end{enumerate}
\end{itemize}

$\judge$ accepts $\accuser$'s claim iff all checks pass.
The verification function $\Verify$ could be represented as 
\begin{equation}
\label{equ:verificationfunc}
    \Verify(F(x_i), y_i):=\mathbb{I}(F(x_i)= y_i~\mathrm{and}~f(x_i)\neq y_i).
\end{equation}


Other model-independent watermarking schemes differ from Adi~\cite{Adi} only in how $(\xx, \yy)$ was sampled; all other parts are the same as Adi~\cite{Adi}.
In Zhang~\cite{zhang}, all samples in $\xx$ are from the same class and perturbed with a secret mask, or they are sampled from a different domain unrelated to the source model's domain.
In Li (a)~\cite{li2019piracy}, all samples in $\xx$ are masked with a small filter consists of three colors: image pixels under the white pattern pixels are changed to a very large negative number, image pixels under black pattern pixels are changed to a very large positive number, and pixels under gray pattern pixels stay unchanged.
Guo~\cite{guo2018watermarking} samples $(\xx, \yy)$ s.t. $f(x_i)\neq y_i$  $\forall (x_i, y_i)\in(\xx,\yy)$, and embeds an $n$-bit message into each $x_i$.
Namba~\cite{Namba} samples $(\xx, \yy)$ s.t. $f(x_i)\neq y_i$  $\forall (x_i, y_i)\in(\xx,\yy)$ and imprints them with greater force and cause the model to learn them profoundly. 
Xu~\cite{xu2019novel} samples $\xx$ but labels them with a serial number.
Therefore, all such schemes follow our generalization.

\subsubsection{Model-dependent: \texorpdfstring{\old{Jia}\new{EWE}~\cite{Jia}}{}}

Recall that $\xx$ sampled in Adi~\cite{Adi} are out-of-distribution.
The watermarked features learnt by $F_\accuser$ are different from the task distribution and can thus be easily removed through compression or other forms of knowledge transfer. 
\old{Jia}\new{EWE}~\cite{Jia} embeds watermarks that are entangled with legitimate data to $F_\accuser$, so that removing such watermarks will sacrifice performance on legitimate data. 
To this end, \old{Jia}\new{EWE} uses the soft nearest neighbor loss (SNNL)~\cite{SNNL1, SNNL2} as an additional loss during training to entangle feature representations of the watermark with the training data.
The data samples contained in $\Proof$ are in-distribution samples from the same class $(\xx, y)$, and these samples will be perturbed by a small mask called trigger.

\begin{itemize}
    \item {\bf claim generation.} $\accuser$ generates $\Proof_\accuser$ as follows:
    \begin{enumerate}
        \item samples in-distribution $(\xx, y)$ s.t. $f(x_i)=y~\forall x_i\in\xx$, for simplicity, we represent it as 
        $f(\xx)= y$;
        \item samples a trigger $t$, computes ${x}'_i:=x_i+t~\forall x_i\in\xx$;\footnote{\revision{The trigger is an input mask which can be arbitrarily chosen by $\accuser$.} And the trigger location is determined as  the area with the largest gradient of SNNL with respect to $x_i$. }
        \item samples ${y}'$ with ${y}'\neq y$;
        \item fine-tunes $F_\accuser$ on $({\xx}', {y}')$;
        \item commits $(\accuser, F_{\accuser}, {\xx}', {y}')$ as $cm_\accuser$ and timestamps.
    \end{enumerate}
    \item {\bf claim verification.} $\judge$ checks the following:
    \begin{enumerate}
        \item 
        $\frac{1}{|\xx'|}\sum\limits_{i=1...|\xx'|}\mathbb{I}(F_\accuser(x'_i)=y'~\mathrm{and}~f(x'_i)\neq y')>T$;
        \item 
        $\frac{1}{|\xx'|}\sum\limits_{i=1...|\xx'|}\mathbb{I}(F_\suspect(x'_i)=y'~\mathrm{and}~f(x'_i)\neq y')>T$;
        \item $cm_{\accuser}$ is a valid commitment of $(\accuser, F_{\accuser}, {\xx}', {y}')$;
           \item $cm_{\accuser}$ was timestamped before $cm_{\suspect}$. 
    \end{enumerate}
\end{itemize}

$\judge$ accepts $\accuser$'s claim iff all checks pass.
The verification function $\Verify$ is also Equation~\ref{equ:verificationfunc}.

\subsubsection{Model-dependent: Li (b)~\texorpdfstring{\cite{ASAC}}{}} 
Li (b) aims to prevent watermark forging by enforcing the trigger set to be indistinguishable from the original samples in the training set. 
To achieve this, it employs an encoder and a discriminator, both trained based on the source model, to construct the trigger set. 
A malicious $\accuser$ who does not have access to the encoder cannot generate a trigger set that works well for the victim model. 
However, as we will show in Section~\ref{sec:li(b)}, it is still possible for $\accuser$ to falsely claim a $F_\suspect$.

\begin{itemize}
    \item {\bf claim generation.} $\accuser$ generates $\Proof_\accuser$ as follows:
    \begin{enumerate}
        \item samples in-distribution $(\xx, \yy)$ s.t. 
        $f(\xx)= \yy$;

        \item generates a triggered $x'_i:~\forall x_i\in\xx$  through the encoder and  discriminator;      
        \item samples ${y_i}'$ with ${y_i}'\neq y_i,~\forall y_i \in \yy$;
        \item fine-tunes $F_\accuser$ on $({\xx}', {\yy}')$; 
        \item commits $(\accuser, F_{\accuser}, {\xx}', {\yy}')$ as $cm_\accuser$ and timestamps.
    \end{enumerate}
    \item {\bf claim verification.} $\judge$ checks the following:
    \begin{enumerate}
        \item 
        $\frac{1}{|\xx'|}\sum\limits_{i=1...|\xx'|}\mathbb{I}(F_\accuser(x'_i)=y_i'~\mathrm{and}~f(x'_i)\neq y_i')>T$;
        \item 
        $\frac{1}{|\xx'|}\sum\limits_{i=1...|\xx'|}\mathbb{I}(F_\suspect(x'_i)=y_i'~\mathrm{and}~f(x'_i)\neq y_i')>T$;
        \item $cm_{\accuser}$ is a valid commitment of $(\accuser, F_{\accuser}, {\xx}', {\yy}')$;
           \item $cm_{\accuser}$ was timestamped before $cm_{\suspect}$. 
    \end{enumerate}
\end{itemize}

$\judge$ accepts $\accuser$'s claim iff all checks pass.
The verification function $\Verify$ is also Equation~\ref{equ:verificationfunc}.

In Frontier-Stitching~\cite{FS}, $(\xx, \yy)$ are generated as adversarial examples w.r.t. $F_\accuser$.
Due to transferability, they are also adversarial examples to the independent models of $F_\accuser$.
Then, $\accuser$ updates $F_\accuser$ with $(\xx, \yy)$ using adversarial training, so that $F_\accuser$ and the models derived from $F_\accuser$ will perform differently from other independent models on $(\xx, \yy)$.
Blackmarks~\cite{BlackMarks} is similar to Frontier-Stitching~\cite{FS} except that all class labels are clustered into two groups; 
the adversarial examples are generated s.t., for randomly selected samples from one cluster, $F_\accuser$ predicts labels from the other cluster.

Both Frontier-Stitching~\cite{FS} and Blackmarks~\cite{BlackMarks} follow our generalization: 
the $\Proof$ includes a trigger set and the claim verification requires running model inference on the trigger set.

\subsubsection{Active: DAWN~\texorpdfstring{\cite{DAWN}}{}}

DAWN is designed to be integrated into the prediction API of a model so that it can survive model extraction. 
It works by dynamically watermarking a small fraction of client queries, altering the model's prediction responses for these queries. 
The watermarked queries act as a trigger set in case an adversarial client attempts to train an extracted model using the prediction responses. 
Using this trigger set, $\accuser$ can demonstrate ownership  as in other DNN watermarking schemes.

\begin{itemize}
    \item {\bf claim generation.} $\accuser$ generates $\Proof_\accuser$ as follows:
    \begin{enumerate}
        \item samples $(\xx, \yy)$ s.t. $F_\accuser(\xx) = \yy$;\footnote{In fact, $\xx$ was received from an API client.}
        \item samples a model-specific key $k$;
        \item for each $x_i\in\xx$, computes  ${y}'_i:=\pi(\mathrm{HMAC}(k, \revision{\mu(x_i)}), y_i)$, \revision{where $\pi()$ is a keyed pseudorandom permutation that that permutes the label from $y_i$ to $y'_i$, and $\mu()$ is a mapping function that ensures $\mu(x)=\mu(x+\delta)$ for a small perturbation $\delta$}\footnote{\revision{Without $\mu()$, a malicious client receiving different predictions for $x$ and $x+\delta$ can discard both $x$ and $x+\delta$ from its training set to avoid the watermark.}};
        \item commits $(\accuser, F_{\accuser}, \xx, \yy, {\yy}')$ as $cm_\accuser$ and timestamps.
    \end{enumerate}
    \item {\bf claim verification.} $\judge$ checks the following:
    \begin{enumerate}
        \item ${y}'_i:=\pi(\mathrm{HMAC}(k, x_i), y_i), \forall x_i\in\xx$;\footnote{The DAWN~\cite{DAWN} paper did not explicitly mention this verification, as robustness against malicious $\accuser$ is not their focus.
        We add this verification because we want to consider the most difficult condition for false claims.}
        \item $\frac{1}{|\xx|}\sum\limits_{i=1...|\xx|}\mathbb{I}(F_\suspect(x_i)=y_i'~\mathrm{and}~f(x_i)\neq y_i')>T$;
        \item $cm_{\accuser}$ is a valid commitment of $(\accuser, F_{\accuser}, \xx, \yy, {\yy}')$;
           \item $cm_{\accuser}$ was timestamped before $cm_{\suspect}$. 
    \end{enumerate}
\end{itemize}

$\judge$ accepts $\accuser$'s claim iff all checks pass.
The verification function $\Verify$ is also Equation~\ref{equ:verificationfunc}.

\subsubsection{Fingerprinting: Lukas~\texorpdfstring{\cite{Lukas}}{}}

Recall that model {fingerprinting} extracts a persistent, identifying code (i.e., fingerprint) from the source model.
Lukas~\cite{Lukas} uses the adversarial examples specific to the source model as the fingerprint.
They hypothesize that there exists a subclass of targeted, transferable, adversarial examples that transfer only to stolen models but not to independent models.
To achieve this, they train a set of independent models and extract a set of surrogate models from the source model.
They then find a set of adversarial examples that minimize the loss for $F_\accuser$ and its extracted models but maximize the loss for independent models.

\begin{itemize}
    \item {\bf claim generation.} $\accuser$ generates $\Proof_\accuser$ as follows:
    \begin{enumerate}
        \item $\accuser$ trains a set of extracted and independent models:
        \begin{itemize}
            \item extracted models are trained on data labeled by $F_\accuser$;
            \item independent models are trained on ground-truth labels.
        \end{itemize}
        \item samples $(\xx,\yy)$ s.t., $f(\xx)\neq\yy$;
        \item for each $x_i\in\xx$, perturb it into ${x}'_i$ s.t., the loss of $({x}'_i, y_i)$ is minimized for $F_\accuser$ and its extracted models, but is maximized for the independent models;
        \item commits $(\accuser, F_{\accuser}, {\xx}', \yy)$ as $cm_\accuser$ and timestamps.
    \end{enumerate}
    \item {\bf claim verification.} $\judge$ checks the following:
    \begin{enumerate}
        \item 
        $\frac{1}{|\xx'|}\sum\limits_{i=1...|\xx'|}\mathbb{I}(F_\accuser(x'_i)=y_i~\mathrm{and}~f(x'_i)\neq y_i)>T$;
        \item 
        $\frac{1}{|\xx'|}\sum\limits_{i=1...|\xx'|}\mathbb{I}(F_\suspect(x'_i)=y_i~\mathrm{and}~f(x'_i)\neq y_i)>T$;
        \item $cm_{\accuser}$ is a valid commitment of $(\accuser, F_{\accuser}, {\xx}', \yy)$;
        \item $cm_{\accuser}$ was timestamped before $cm_{\suspect}$.  
    \end{enumerate}
\end{itemize}

$\judge$ accepts $\accuser$'s claim iff all checks pass. 
The verification function $\Verify$ is also Equation~\ref{equ:verificationfunc}.

\old{Zhao}\new{AFA}~\cite{Zhao} is a concurrent and independent work with Lukas~\cite{Lukas}: 
it also generates the adversarial examples specific to the source model.
IPGuard~\cite{IPGuard} is similar to Lukas~\cite{Lukas} except that it uses a different way to find adversarial examples {\em near} the decision boundary of $F_\accuser$:
they start from an initial data point and iteratively move it along the gradient of the objective function. 
Metav~\cite{Metav} is an extension of Lukas~\cite{Lukas}:
besides generating $(\xx', \yy)$ in the same way as in~\cite{Lukas}, it also trains a verifier;
the concatenated outputs of $F_\suspect$ on $(\xx', \yy)$ will be input to the verifier, which then determines if $F_\suspect$ is a stolen model.
Therefore, both IPGuard~\cite{IPGuard} and Metav~\cite{Metav} follow our generalization.

\subsubsection{Fingerprinting: DI~\texorpdfstring{\cite{DI}}{}}

The key observation for DI is that all models derived from $F_\accuser$ will contain direct or indirect information from the training set of $F_\accuser$,
hence model ownership can be resolved by showing that the suspect model was trained (at least partially or indirectly) on the same dataset as the source model.
DI measures $F_\suspect$'s prediction margins (distances from the samples to the model's decision boundaries) for both $F_\accuser$'s training set and public samples. 
If  $F_\suspect$ has different prediction margins for them, it is deemed to be stolen; otherwise the model is deemed independent.

\begin{itemize}
    \item {\bf claim generation.} $\accuser$ generates $\Proof_\accuser$ as follows:
    \begin{enumerate}
        \item samples $\xx$ from the training dataset of $F_\accuser$; extracts the feature embeddings for $\xx$ that characterizes its ``prediction margin'' (i.e., distance from the decision boundaries) w.r.t. $F_\accuser$; and labels them as ``inside'' ($b=1$); 
        \item samples $\xx'$ from an unseen publicly available dataset; extracts the feature embeddings for $\xx'$ w.r.t. $F_\accuser$; and labels them as ``outside'' ($b=0$); 
        \item using the embeddings and the ground truth membership labels (i.e., $b$), trains a regression model $g_{\accuser}$, which is to predict a (proxy) measure of confidence that a sample is in the training set of $F_\accuser$.
        \item commits $(\accuser, F_{\accuser}, (\xx,1), (\xx',0), g_\accuser)$ as $cm_\accuser$ and timestamps.
    \end{enumerate}
    \item {\bf claim verification.} $\judge$ checks the following:
    \begin{enumerate}
        \item extracts the feature embeddings for $\xx$ and $\xx'$, w.r.t. $F_\accuser$; inputs the embeddings to $g_\accuser$,  gets the confidence scores, and checks if there is a significant difference between the two sets of confidence scores;
        \item extracts the feature embeddings for $\xx$ and $\xx'$, w.r.t. $F_\suspect$; inputs the embeddings to $g_\accuser$, gets the confidence scores, and checks if there is a significant difference between the two sets of confidence scores;
        \item $cm_{\accuser}$ is a valid commitment of $(\accuser, F_{\accuser}, (\xx,1), (\xx',0), g_\accuser)$;
           \item $cm_{\accuser}$ was timestamped before $cm_{\suspect}$. 
    \end{enumerate}
\end{itemize}

$\judge$ accepts $\accuser$'s claim iff all checks pass.
DI still follows our generation as it requires running a model on the ``trigger set'' to generate embeddings. 
However, it differs from other MOR schemes in that it does not verify each sample separately; 
instead, it verifies the trigger set as a whole,
calculating the effect size or $p$-value between the confidence scores of $\xx$ and $\xx'$, and comparing it  with a threshold $T$.


\section{Transferable Adversarial Examples Against MOR}
\label{sec:attack}

Recall that most of the samples in the trigger set $(\xx, \yy)$ satisfy: $\Verify(F_{\accuser}(x_i), y_i)=1$ for $F_{\accuser}$ and $\Verify(F(x_i),y_i)=0$ for other independently trained $F\neq F_{\accuser}$.
To falsely claim the ownership of an independent model $F$, $\accuser$ could just generate $(x_i,y_i)$ in a way s.t. $\Verify(F(x_i), y_i)=1$ for most $(x_i, y_i)\in(\xx, \yy)$.
Our key observation is that $\accuser$ can achieve this by \ul{leveraging the transferability of adversarial examples}. 
Recall that we assume $\accuser$ has access to a dataset that has the same distribution with $F_\suspect$'s training set.
Then, we could have $\accuser$ train $F_\accuser$ using this dataset and generate a malicious trigger set $\hat{\xx}$ as transferable adversarial examples for $F_\accuser$. 
Then, $\Verify(F(\hat{x}_i), y_i)=1$ for most $(\hat{x}_i, y_i)\in(\hat{\xx}, \yy)$, holds for the independent models of $F_{\accuser}$.


\subsection{Attacks in detail}

For each exemplary scheme from Section~\ref{sec:survey}, we show how a malicious $\accuser$ generates $\Proof_\accuser$ so that it can falsely claim ownership of an independent model.
The \ul{underlined} text describes $\accuser$'s misbehaviour. 
For all attacks, the training sets of $F_\accuser$ and $F_\suspect$ are distinct but follow the same distribution.

\subsubsection{Model-independent: Adi~\texorpdfstring{\cite{Adi}}{}}

Recall that Adi~\cite{Adi} fine-tunes $F_\accuser$ on $(\xx, \yy)$, where $f(x_i)\neq y_i$  $\forall (x_i, y_i)\in(\xx,\yy)$; models derived from $F_\accuser$ have good performance on $(\xx, \yy)$.
$\accuser$ could generate $(\xx, \yy)$ as adversarial examples so that $F_\accuser$ performs well on $(\xx, \yy)$.
Due to transferability, models that are independent of $F_\accuser$ 
also perform well on $(\xx, \yy)$.
Then, $\accuser$ can claim ownership of these models.

\begin{itemize}
    \item {\bf claim generation.} $\accuser$ generates $\Proof_\accuser$ as follows:
    \begin{enumerate}
        \item samples $\xx$ and assigns a wrong label $y_i$ to each $x_i \in \xx$, i.e., $f(\xx)\neq \yy$;
        \item \ul{for each $x_i\in\xx$, perturbs it into $\hat{x}_i$ s.t., $F_\accuser(\hat{\xx})=\yy$}. 
        \item commits \ul{$(\accuser, F_{\accuser}, \hat{\xx}, \yy)$} as $cm_\accuser$ and timestamps.
    \end{enumerate}
    \revision{
    \item {\bf claim verification.} $\judge$ checks the following:
    \begin{enumerate}
        \item 
        $\frac{1}{|\hat{\xx}|}\sum\limits_{i=1...|\hat{\xx}|}\mathbb{I}(F_\accuser(\hat{x}_i)=y_i~\mathrm{and}~f(\hat{x}_i)\neq y_i)>T$;
        \item 
        $\frac{1}{|\hat{\xx}|}\sum\limits_{i=1...|\hat{\xx}|}\mathbb{I}(F_\suspect(\hat{x}_i)=y_i~\mathrm{and}~f(\hat{x}_i)\neq y_i)>T$;
        \item $cm_{\accuser}$ is a valid commitment of $(\accuser, F_{\accuser}, \hat{\xx}, {\yy})$;
        \item $cm_{\accuser}$ was timestamped before $cm_{\suspect}$. 
    \end{enumerate}
    The first check holds because $(\hat{\xx}, \yy)$ are adversarial examples generated for $F_\accuser$.
    The second check holds because $(\hat{\xx}, \yy)$ can transfer to $F_\suspect$.
    }
\end{itemize}

In fact, 
the labels $\yy$ can be chosen based on the perturbation of $\xx$, i.e., untargeted adversarial examples.
This makes the adversarial optimization easy to converge. 

Other model-independent watermarking schemes  differ from Adi~\cite{Adi} only in how $(\xx, \yy)$ was sampled, hence  false claims can be done in the same way as in Adi. 

\subsubsection{Model-dependent:~\texorpdfstring{ \old{Jia}\new{EWE}~\cite{Jia}}{}}

False claims for \old{Jia}\new{EWE}~\cite{Jia} are similar to Adi~\cite{Adi}, except that $\accuser$ needs to generate targeted adversarial examples this time.
Specifically, in \old{Jia}\new{EWE}~\cite{Jia}, all samples in $\xx'$ have the same label $y'$.
Then, $\accuser$ can no longer determine the label based on the perturbation of $\xx$.
Instead, it has to perturb each $x_i$ for a target label.
Targeted adversarial examples can be generated in the same way, but its transferability becomes weaker. 
In Section~\ref{sec:enhancement}, we describe a way for enhancing its transferability.

\begin{itemize}
    \item {\bf claim generation.} $\accuser$ generates $\Proof_\accuser$ as follows:
    \begin{enumerate}
        \item samples $(\xx, y)$ with $f(\xx)= y$;
        \item samples a trigger $t$, computes ${x}'_i:=x_i+t,~\forall x_i\in\xx$;
        \item samples ${y}'$ with ${y}'\neq y$;
        \item \ul{for each $x'_i\in\xx'$, perturbs it into $\hat{x}_i$ s.t., $F_\accuser(\hat{\xx})=y'$}. 


        \item commits \ul{$(\accuser, F_{\accuser}, \hat{\xx}, {y}')$} as $cm_\accuser$ and timestamps.
    \end{enumerate}
    \revision{
    \item {\bf claim verification.} $\judge$ checks the following:
    \begin{enumerate}
        \item 
        $\frac{1}{|\hat{\xx}|}\sum\limits_{i=1...|\hat{\xx}|}\mathbb{I}(F_\accuser(\hat{x}_i)=y'~\mathrm{and}~f(\hat{x}_i)\neq y')>T$;
        \item 
        $\frac{1}{|\hat{\xx}|}\sum\limits_{i=1...|\hat{\xx}|}\mathbb{I}(F_\suspect(\hat{x}_i)=y'~\mathrm{and}~f(\hat{x}_i)\neq y')>T$;
        \item $cm_{\accuser}$ is a valid commitment of $(\accuser, F_{\accuser}, \hat{\xx}, {y}')$;
           \item $cm_{\accuser}$ was timestamped before $cm_{\suspect}$. 
    \end{enumerate}
    The first check holds because $(\hat{\xx}, y')$ are adversarial examples generated for $F_\accuser$.
    The second check holds because $(\hat{\xx}, y')$ can transfer to $F_\suspect$.
    }
\end{itemize}

\subsubsection{Model-dependent: Li (b)~\texorpdfstring{\cite{ASAC}}{}}
\label{sec:li(b)}
Recall that Li (b) claims that a malicious $\accuser$ who does not have access to the encoder cannot generate a valid trigger set.
However, this claim does not take into account the possibility of transferable adversarial examples.
That is, a malicious $\accuser$ without knowing the encoder can still generate a set of transferable adversarial examples as the trigger set, with the condition that the adversarial examples are within an $L_\epsilon$ neighborhood of the original samples to satisfy the indistinguishability requirement between the trigger set and its original samples.


\begin{itemize}
\item {\bf claim generation.} $\accuser$ generates $\Proof_\accuser$ as follows:
    \begin{enumerate}
        \item samples in-distribution $(\xx, \yy)$ s.t. 
        $f(\xx)= \yy$;

        \item generates a triggered $x'_i:~\forall x_i\in\xx$  through the encoder and  discriminator;      
        \item samples ${y_i}'$ with ${y_i}'\neq y_i,~\forall y_i \in \yy$;
         \item \ul{for each $x'_i\in\xx'$, perturbs it into $\hat{x}_i$ s.t., $F_\accuser(\hat{\xx})= \yy'$}. 
        \item commits $(\accuser, F_{\accuser}, \hat{\xx}, {\yy}')$ as $cm_\accuser$ and timestamps.
    \end{enumerate}
        \revision{
    \item {\bf claim verification.} $\judge$ checks the following:
    \begin{enumerate}
        \item 
        $\frac{1}{|\hat{\xx}|}\sum\limits_{i=1...|\hat{\xx}|}\mathbb{I}(F_\accuser(\hat{x}_i)=y'_i~\mathrm{and}~f(\hat{x}_i)\neq y'_i)>T$;
        \item 
        $\frac{1}{|\hat{\xx}|}\sum\limits_{i=1...|\hat{\xx}|}\mathbb{I}(F_\suspect(\hat{x}_i)=y'_i~\mathrm{and}~f(\hat{x}_i)\neq y'_i)>T$;
        \item $cm_{\accuser}$ is a valid commitment of $(\accuser, F_{\accuser}, \hat{\xx}, {\yy'})$;
        \item $cm_{\accuser}$ was timestamped before $cm_{\suspect}$. 
    \end{enumerate}
    The first check holds because $(\hat{\xx}, \yy')$ are adversarial examples generated for $F_\accuser$.
    The second check holds because $(\hat{\xx}, \yy')$ can transfer to $F_\suspect$.
    }
\end{itemize}

False claims for Frontier-Stitching~\cite{FS} and Blackmarks~\cite{BlackMarks} are easy:
$\accuser$ can simply generate $(\xx, \yy)$ s.t. $f(\xx)=\yy$; then all independent models will behave the same as $F_\accuser$.

\subsubsection{Active: DAWN~\texorpdfstring{\cite{DAWN}}{}}

Similar to \old{Jia}\new{EWE}~\cite{Jia}, $\accuser$ again needs to generate targeted adversarial examples to attack DAWN~\cite{DAWN}.

\begin{itemize}
    \item {\bf claim generation.} $\accuser$ generates $\Proof_\accuser$ as follows:
    \begin{enumerate}
        \item samples $(\xx, \yy)$ s.t. $F_\accuser(\xx) = \yy$;
        \item samples a model-specific key $k$;
        \item for each $x_i\in\xx$, computes  ${y}'_i:=\pi(\mathrm{HMAC}(k, \revision{\mu{(x_i)}}), y_i)$;
        \item \ul{for each $x_i\in\xx$, perturbs it into $\hat{x}_i$ s.t., $F_\accuser(\hat{\xx})=\yy'$}.\footnote{In principle, $\accuser$ cannot modify $\xx$ as it is supposed to be received from an API client. However, without some additional means of validation, $\accuser$ can simply claim that $\hat{\xx}$ was received from an API client (cf. Section~\ref{sec:countermeasures}).} 
        \item commits \ul{$(\accuser, F_{\accuser}, \hat{\xx}, \yy, {\yy}')$} as $cm_\accuser$ and timestamps.
    \end{enumerate}
    \revision{
        \item {\bf claim verification.} $\judge$ checks the following:
    \begin{enumerate}
        \item ${y}'_i:=\pi(\mathrm{HMAC}(k, \mu(\hat{x}_i)), y_i), \forall x_i\in\hat{\xx}$;
        \item $\frac{1}{|\hat{\xx}|}\sum\limits_{i=1...|\hat{\xx}|}\mathbb{I}(F_\suspect(\hat{x}_i)=y_i'~\mathrm{and}~f(\hat{x}_i)\neq y_i')>T$;
        \item $cm_{\accuser}$ is a valid commitment of $(\accuser, F_{\accuser}, \hat{\xx}, \yy, {\yy}')$;
        \item $cm_{\accuser}$ was timestamped before $cm_{\suspect}$. 
    \end{enumerate}
    The first check holds because     $\mu(x_i)=\mu(\hat{x}_i)$\footnote{\revision{
    In our experiments, we took $\mu()$ from the open-sourced implementation of DAWN and it shows that more than 88\% of the adversarial examples satisfy $\mu(x_i)=\mu(\hat{x}_i)$. Furthermore, $\accuser$ can discard the $\hat{x}_i$s that do not satisfy this condition.}}.
    The second check holds because $(\hat{\xx}, \yy')$ can transfer to $F_\suspect$. 
    }
\end{itemize}

Recall that DAWN changes the prediction results of $F_\accuser$ instead of $F_\accuser$ itself.
Therefore, a source model $F_\accuser$, trained by an honest accuser, should in principle have a low performance on the committed samples. 
Based on this observation, we could prevent the above attack by introducing an additional check during verification:
$\judge$ checks the performance of $F_\accuser$ on the committed samples.
However, this again can be attacked.
Namely, after Step~4, a malicious $\accuser$ could update $F_\accuser$ with $(\hat{\xx}, {\yy})$ in a way like adversarial training.

\subsubsection{Fingerprinting: Lukas~\texorpdfstring{\cite{Lukas}}{}}

Recall that Lukas~\cite{Lukas} uses adversarial examples specific to the source model as the fingerprint:
they train a set of extracted models and independent models, and perturb the adversarial examples to minimize the loss for the extracted models but maximize the loss for independent models.
We could simply omit the part of maximizing loss for independent models, then the adversarial examples will transfer to  independent models.

\begin{itemize}
    \item {\bf claim generation.} $\accuser$ generates $\Proof_\accuser$ as follows:
    \begin{enumerate}
        \item samples $(\xx,\yy)$ s.t., $f(\xx)\neq\yy$;
        \item \ul{for each $x_i\in\xx$, perturbs it into $\hat{x}_i$ s.t., $F_\accuser(\hat{\xx})=\yy$;}
        \item commits $(\accuser, F_{\accuser}, \hat{\xx}, \yy)$ as $cm_\accuser$ and timestamps.
    \end{enumerate}
    \revision{
        \item {\bf claim verification.} $\judge$ checks the following:
    \begin{enumerate}
        \item 
        $\frac{1}{|\hat{\xx}|}\sum\limits_{i=1...|\hat{\xx}|}\mathbb{I}(F_\accuser(\hat{x}_i)=y_i~\mathrm{and}~f(\hat{x}_i)\neq y_i)>T$;
        \item 
        $\frac{1}{|\hat{\xx}|}\sum\limits_{i=1...|\hat{\xx}|}\mathbb{I}(F_\suspect(\hat{x}_i)=y_i~\mathrm{and}~f(\hat{x}_i)\neq y_i)>T$;
        \item $cm_{\accuser}$ is a valid commitment of $(\accuser, F_{\accuser}, \hat{\xx}, \yy)$;
        \item $cm_{\accuser}$ was timestamped before $cm_{\suspect}$.  
    \end{enumerate}
    The first check holds because $(\hat{\xx}, \yy)$ are adversarial examples generated for $F_\accuser$.
    The second check holds because $(\hat{\xx}, \yy)$ can transfer to $F_\suspect$.
    }
\end{itemize}

Again, $\accuser$ can choose $\yy$ based on the perturbation of $\xx$, as untargeted adversarial examples.

\old{Zhao}\new{AFA}~\cite{Zhao}, IPGuard~\cite{IPGuard} and Metav~\cite{Metav} can be attacked in the same way as Lukas~\cite{Lukas}, except that in Metav~\cite{Metav} the adversarial optimization needs to be based on the verifier.

\subsubsection{Fingerprinting: DI~\texorpdfstring{\cite{DI}}{}}

Recall that $\accuser$ in DI needs to include a regression model $g_\accuser$ in $\Proof_\accuser$.
Therefore, to falsely claim a model in DI~\cite{DI}, $\accuser$ could simply manipulate $g_\accuser$ s.t. it always shows significant differences even for independent $F_\suspect$.
Recall that we assume $\accuser$ has access to a dataset with the same distribution as $F_\suspect$'s training set.
We could have $\accuser$ train $F_\accuser$ with this dataset and ``adversarially'' perturb each sample in $\xx$ with a small amount of noise such that the ``prediction margin'' (distance between $\xx$ and the decision boundaries) w.r.t. $F_\accuser$ are larger.
Due to the transferability of our adversarially perturbed samples, $F_\suspect$ will also have a larger ``prediction margin'' on $\xx$.
Then if $g_\accuser$ was trained with the embeddings of this perturbed $\xx$ and public $\xx'$ w.r.t $F_\accuser$, it will output significantly different confidence scores when it takes the embeddings of $\xx$ and $\xx'$ w.r.t. $F_\suspect$.


\begin{itemize}
    \item {\bf claim generation.} $\accuser$ generates $\Proof_\accuser$ as follows:
    \begin{enumerate}
        \item 
        samples $\xx$ from the training dataset of $F_\accuser$;
        \ul{perturbs $\xx$ s.t. 
        the ``prediction margin'' w.r.t. $F_\accuser$ are larger};
        extracts the feature embeddings for $\xx$ that characterizes this ``prediction margin''; and labels them as ``inside'' ($b=1$); 
        \item samples $\xx'$ from an unseen publicly available dataset; extracts the feature embeddings for $\xx'$ w.r.t. $F_\accuser$; and labels them as ``outside'' ($b=0$); 
        \item using the embeddings and the ground truth membership labels (i.e., $b$), trains a regression model $g_{\accuser}$, which is to predict a (proxy) measure of confidence that a sample is in the training set of $F_\accuser$;
        \item commits {$(\accuser, F_{\accuser}, ({\xx},1), (\xx',0), g_\accuser)$} as $cm_\accuser$ and timestamps.
    \end{enumerate}
    \revision{
    \item {\bf claim verification.} $\judge$ checks the following:
    \begin{enumerate}
        \item extracts the feature embeddings for $\xx$ and $\xx'$, w.r.t. $F_\accuser$; inputs the embeddings to $g_\accuser$,  gets the confidence scores, and checks if there is a significant difference between the two sets of confidence scores;
        \item extracts the feature embeddings for $\xx$ and $\xx'$, w.r.t. $F_\suspect$; inputs the embeddings to $g_\accuser$, gets the confidence scores, and checks if there is a significant difference between the two sets of confidence scores;
        \item $cm_{\accuser}$ is a valid commitment of $(\accuser, F_{\accuser}, (\xx,1), (\xx',0), g_\accuser)$;
           \item $cm_{\accuser}$ was timestamped before $cm_{\suspect}$. 
    \end{enumerate}
    The first check holds because $\xx$ are ``adversarially'' perturbed w.r.t. $F_\accuser$.
    The second check holds because $\xx$ can transfer to $F_\suspect$.
    }
\end{itemize}

\subsection{Transferability enhancement}
\label{sec:enhancement}

We borrow an idea of Lukas~\cite{Lukas} to enhance the transferability.
Recall that $\accuser$ aims to generate $(\hat{x},y)$ in a way s.t. $\Verify(F(\hat{x}), y)=1$ for all independent models of $F_\accuser$.
We could have $\accuser$ generate a set of independent models, and perturb the adversarial examples to minimize the loss of these independent models for $\Verify(F(\hat{x}),y) = 1$. 

To generate an untargeted adversarial example, $\accuser$ just needs to {\em maximize} the following loss function:
\begin{equation}
\label{equ:loss}
\begin{aligned}
    \mathcal{L}(\hat{x}, y)  = & L(F_{\accuser}(\hat{x}),y) + \sum_{F\in {\mathcal{F}}}\beta_F L(F(\hat{x}),y),
\end{aligned}
\end{equation} 
where $y$ is the true label, $L()$ denotes the cross-entropy loss, 
$\mathcal{F}$ denotes a set of independent models, 
$\beta_F$ is a weight \footnote{We set $\beta_F = \frac{1}{|F|}$ in our experiments.} for $F$.
We use Iterative Fast Gradient Sign Method (IFGSM)~\cite{IFGSM} to find a solution that both 
maximizes $\mathcal{L}(\hat{x}, y)$ and
minimizes the perturbation: 
\begin{equation*}
    \begin{aligned}
    \hat{x} := Clip_{\hat{x}, \epsilon}\{\hat{x}+\alpha sign(\Delta_{\hat{x}}\mathcal{L}(\hat{x}, y))\},
    \end{aligned}
\end{equation*}
where $Clip_{\hat{x},\epsilon}()$ is a function that performs per-pixel clipping for $\hat{x}$. We use $\alpha=0.03$ and select the number of iteration to be 100 to make the adversarial examples reach the edge of the $\epsilon$ max-norm ball.
It ensures that the generated adversarial example $\hat{x}$ will be in $L_{\infty} \epsilon$-neighborhood of the original input $x$. 
Notice that the per-pixel perturbation bound $\epsilon$ balances the false claim effectiveness and the perturbation visibility.

To generate a targeted adversarial example, $\accuser$ needs to {\em minimize} the loss function  (i.e., Equation~\ref{equ:loss}), where $y$ becomes the target (wrong) label.
The iteration step is similar to that in the untargeted case:
\begin{equation*}
    \begin{aligned}
    \hat{x} := Clip_{\hat{x}, \epsilon}\{\hat{x}-\alpha sign(\Delta_{\hat{x}}\mathcal{L}(\hat{x}, y))\}.
    \end{aligned}
\end{equation*}

This enhancement is generally applicable and we have implemented it for all of our benchmarks in Section~\ref{sec:eval}.
We remark that our main contribution is to show that defending against false claims is as challenging as preventing transferable adversarial examples. 
Any method for generating transferable adversarial examples is pluggable into our attacks. 
There are many methods for transferability enhancement; it is only necessary to show that our attack works with one such method.

\begin{table*}[ht]
\caption{\revision{Decision thresholds (we refer to Section~\ref{sec:thresholds} for an explanation of the decision thresholds; 
the column of DI shows the normalized effect size $(\%)$ instead of $\mathit{MORacc}$
\label{tab:decision_thresholds})}.
}
\centering
\begin{spacing}{1.30}
\begin{tabular}{c|c|c|c|c|c|c|c}
\toprule[1pt]
Dataset&Threshold type&Adi &\old{Jia}\new{EWE} &Li (b) &DAWN &Lukas &DI \\ \toprule[0.8pt]

\multicolumn{1}{c|}{\multirow{3}{*}{CIFAR-10}} 
&independent & 10.0  & 1.8  & 23.0   & 1.0 & 28.0 & 90.0  \\\cline{2-8}
 &  mixed  & 29.0 & 32.9  & 61.5 & 38.5 & 57.5 & 81.4 \\\cline{2-8}
 & extracted & 48.0 &64.0& 100.0 & 76.0 & 87.0 & 72.8 \\ \cline{1-8}

\multicolumn{1}{c|}{\multirow{3}{*}{ImageNet}} 
&independent & 15.0   & 12.0  & 30.0   & 3.0 & 14.0  &  76.5  \\\cline{2-8}
& mixed  &23.5  & 37.5   & 65.0   & 42.5 & 30.0 &  69.6 \\\cline{2-8}
& extracted & 32.0 & 63.0   & 100.0   & 82.0 & 46.0 & 62.6  \\ \cline{1-8}

\multicolumn{1}{c|}{\multirow{3}{*}{CelebA}} 
&independent &25.7&3.7&55.0&7.0&21.0& 20.0  \\\cline{2-8}                     
& mixed &42.4&2.9&55.5&26.0&28.5&14.1  \\\cline{2-8}
& extracted &59.0&2.0&56.0&45.0&36.0&8.2  \\ 
\toprule[0.8pt]
\end{tabular}
\end{spacing}
\end{table*}

\begin{table*}[ht]
\caption{\revision{False claim effectiveness (the bold number means $\mathit{MORacc}$ is higher than the mixed threshold and the underlined number means $\mathit{MORacc}$ is higher than the ``extracted'' threshold
\label{tab:falseclaimeffectiveness}).}
}
\centering
\begin{spacing}{1.30}
\begin{tabular}{c|c|c|c|c|c|c|c}
\toprule[1pt]
Dataset&Model configurations ($F_\suspect$ vs. $F_\accuser$) &Adi &\old{Jia}\new{EWE} &Li (b) &DAWN &Lukas &DI \\ \toprule[0.8pt]

\multicolumn{1}{c|}{\multirow{3}{*}{CIFAR-10}} & 
different structures \& different data &\bu{94.3} & \bu{69.3}  &  \bo{94.3}  & \bo{69.3} & \bu{94.3} & \bu{100} \\\cline{2-8}
&same structure \& different data &\bu{98.0} & \bu{100.0} & \bo{98.0}  & \bu{100.0} & \bu{98.0} & \bu{99.1} \\\cline{2-8}
&different structures \& same data &\bu{99.0} & \bu{78.3}  & \bo{99.0} &  \bu{78.3} & \bu{99.0} & \bu{98.6} \\ \cline{1-8}

\multicolumn{1}{c|}{\multirow{3}{*}{ImageNet}} 
&different structures \& different data  &\bu{72.6} & \bu{87.6}  & \bo{72.6}  &\bu{87.6}   &\bu{72.6} & \bu{100}\\\cline{2-8}
&same structure \& different data &\bu{93.7}    & \bu{97.0}  & \bo{93.7}  & \bu{97.0}  &\bu{93.7}  & \bu{100} \\\cline{2-8}
&different structures \& same data & \bu{84.6}    & \bu{89.0}  & \bo{84.6}  & \bu{89.0}  &\bu{84.6}  & \bu{100}\\ \hline

\multicolumn{1}{c|}{\multirow{1}{*}{CelebA}} 
& \multicolumn{1}{c|}{\begin{tabular}[c]{@{}c@{}}different structures \& different data\\ (Amazon Rekognition API)\end{tabular}} &\bu{68.4}&\bu{68.0}&\bu{68.4}&\bu{68.0}&\bu{68.4}& \bu{99.9} \\

\toprule[0.8pt]
\end{tabular}
\end{spacing}
\end{table*}

\section{Evaluation}
\label{sec:eval}

We now empirically evaluate our attacks against Adi~\cite{Adi}, \old{Jia}\new{EWE}~\cite{Jia}, Li (b)~\cite{ASAC}, DAWN~\cite{DAWN}, Lukas~\cite{Lukas} and DI~\cite{DI}.

\subsection{Setup}

For Adi and \old{Jia}\new{EWE}, we use the Watermark-Robustness-ToolBox\footnote{\url{https://github.com/dnn-security/Watermark-Robustness-Toolbox}} to reproduce their results; 
for Li (b)\footnote{\url{https://github.com/zhenglisec/Blind-Watermark-for-DNN}}, DAWN\footnote{{https://github.com/ssg-research/dawn-dynamic-adversarial-watermarking-of-neural-networks}}
and DI\footnote{\url{https://github.com/cleverhans-lab/dataset-inference}}, we use their open-sourced implementations;
as the source code of Lukas\footnote{\url{https://github.com/ayberkuckun/DNN-Fingerprinting}} is based on TensorFlow (whereas all others are in Pytorch), we re-implemented their scheme in Pytorch~\cite{Paszke2019PyTorch}. 

We consider all three kinds of decision thresholds we discussed in Section~\ref{sec:thresholds}.
For the extracted models, we use fine-tuning based extraction, a.k.a. Fine-Tune All layers (FTAL)~\cite{Uchida}\footnote{
\revision{FTAL comprises two steps:
it first queries $F_{\accuser}$ with a  set of samples and obtains the predicted labels;
it then uses this set to fine-tune $F_{\accuser}$ with a smaller learning rate.
The model extracted through FTAL is very similar to $F_{\accuser}$, 
hence it provides the most challenging threshold for false claims. Indeed, 
}FTAL is the most effective one among several model extraction methods we have tried.
}.\old{,
i.e., using a training set labeled by $F_{\accuser}$ to fine-tune the extracted models.}

For datasets, we consider CIFAR-10~\cite{CIFAR10}, ImageNet~\cite{ImageNet}, and a face attributes dataset named CelebA~\cite{liu2015faceattributes}.

\revision{
When we measure the $\mathit{MORacc}$ on CIFAR-10 and ImageNet, we consider the following three cases:
\begin{enumerate}
    \item {``different structures \& different data'':} $F_{\accuser}$ and $F_{\suspect}$ use different model structures and training data.
    \begin{itemize}
        \item We divided CIFAR-10 into two non-overlapping subsets to train $F_\accuser$ and $F_\suspect$ respectively.
        The model structure for $F_\accuser$ is ResNet 28$\times$10~\cite{wideResnet} and it achieves an accuracy of 89.3\%; the model structure for $F_\suspect$ is ResNet-34~\cite{Resnet} and it achieves an accuracy of 86.3\%.
        \item We randomly selected ten classes from ImageNet to form an ImageNet-10 dataset, and divided ImageNet-10 into two non-overlapping subsets to train $F_\accuser$ and $F_\suspect$ respectively. The model structure for $F_\accuser$ is ResNet-18~\cite{Resnet} and it achieves an accuracy of 82.0\%;
    we use VGG-13~\cite{VGG} for $F_\suspect$ and it achieves an accuracy of 85.9\%.
    \end{itemize}
    \item {``same structure \& different data'':} $F_{\accuser}$ and $F_{\suspect}$ use the same model structure (ResNet 28$\times$ 10 for CIFAR-10 and ResNet-18 for ImageNet), but different training data (as described in case~1).
    \item {``different structures \& same data'':}
    $F_{\accuser}$ and $F_{\suspect}$ use the same training data, butdifferent model structures (as described in case~1).
\end{enumerate}
We aim to use CelebA to train a model (i.e., $F_{\accuser}$) to falsely claim the model behind Amazon Rekognition API~\cite{Amazon}, which detects face features of uploaded images.
The model structure for $F_\accuser$ is Resnet-18~\mbox{\cite{Resnet}} and it achieves an accuracy of  97.8\%.
We have no knowledge about the training dataset or the model architecture of Amazon Rekognition, except that we have black-box access to the model via the API.
Therefore, the setting for CelebA is ``different structures \& different data''.
}

For all three datasets, we set the trigger set size as 100.
$F_\accuser$'s $\mathit{MORacc}$ on the trigger sets are near 100\% for all MOR schemes except DAWN, which has a  $\mathit{MORacc}$ of 0 because its $F_\accuser$ is independent of its trigger sets.
For all experiments, we set the per-pixel perturbation bound, (i.e., maximal value allowed for per-pixel perturbation) as 16.

\revision{
Recall that we use independent models to enhance transferability (cf. Section~\ref{sec:enhancement}).
We train 10 independent models for CIFAR-10, 
8 independent models for ImageNet, 
and 2 independent models for CelebA.\footnote{\revision{We began with two independent models in each case. With CelebA, two models were enough to get  high $\mathit{MORacc}$.
On CIFAR-10 and ImageNet, the $\mathit{MORacc}$ with two models were not high enough, hence we tried increasing the number of independent models until a sufficiently high $\mathit{MORacc}$ was reached: 10 for CIFAR-10 and 8 for ImageNet. However, training models on CIFAR-10 and ImageNet is not as expensive as in CelebA.}
}
}

All reported results are averages of 5 runs; all reported run times  were obtained using (single) Tesla V100 GPUs.
We refer to Appendix~\ref{sec:parameters} for more details of our experiments.

\subsection{False claim effectiveness}

Table~\ref{tab:decision_thresholds} presents the decision thresholds and Table~\ref{tab:falseclaimeffectiveness} presents $F_\suspect$'s $\mathit{MORacc}$ for all three datasets.
It demonstrates that, \ul{for a realistic configuration of MORs using the mixed threshold, our false claims can succeed against all MOR schemes on all datasets}.
Even for the extracted threshold (least favorable to a malicious $\accuser$),  false claims still succeed against all except Li (b).
We also show that in scenarios where $F_\accuser$ and $F_\suspect$ are trained with either the same model structure  or the same dataset, false claims become much easier.

Recall that the $\Proof$-generation procedure for Adi, \old{Jia}\new{EWE}, Li (b), DAWN and Lukas are similar: 
generating adversarial examples that transfer to $F_\suspect$.
The untargeted adversarial examples generated by Adi, Li~(b), and Lukas result in identical $\mathit{MORacc}$. 
In contrast, \old{Jia}\new{EWE} and DAWN's targeted adversarial examples yield a different $\mathit{MORacc}$ value.
The distinguishing factor lies in the decision thresholds of each scheme.

Recall that DI calculates the effect size or $p$-value between the confidence scores of training samples and public samples, and compares it with a threshold, hence the last column in Table~\ref{tab:decision_thresholds}~and ~\ref{tab:falseclaimeffectiveness} shows the normalized effect size instead of $\mathit{MORacc}$
(using normalized effect sizes is equivalent to using the $p$-values directly). 
Different from other MOR schemes, DI treats $F_\suspect$ as stolen as long as $F_\suspect$ was trained with the same data as $F_\accuser$, even if it was trained independently.
It is possible that the model behind Amazon Rekognition API was trained from CelebA as well, in which case it is {not} independent from our $F_\accuser$ in terms of DI.
Therefore, we tested DI (without our attack) on Amazon Rekognition API and obtained a normalized effect size of $0.0$ indicating that there was likely no overlap between CelebA and the training data used for Rekognition API. 

\begin{figure*}[ht]
    \centering
    \subfigure[CIFAR-10]{
    \includegraphics[width=.34\linewidth]{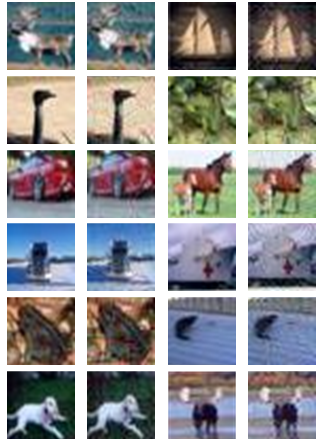}
    }
    \subfigure[ImageNet]{
    \includegraphics[width=.34\linewidth]{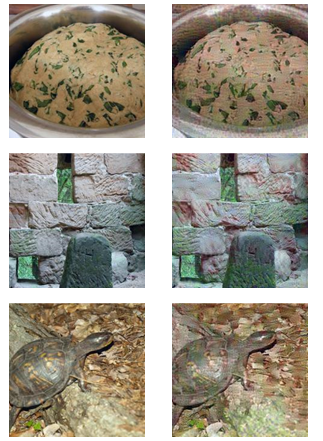}
    }
    \subfigure[CelebA]{
    \includegraphics[width=.26\linewidth]{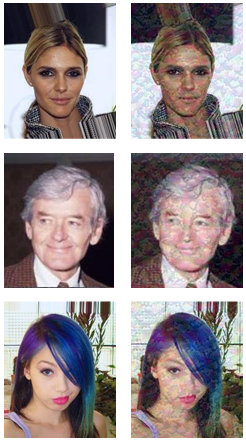}
    }  
    \caption{Comparison between original images and the noised versions (the per-pixel perturbation bound is 16; the original images are on the left-hand side and the noised images are on the right-hand side).}
    \label{fig:comp_pic}
\end{figure*}

\fig~\ref{fig:comp_pic} shows a comparison between an original image in the trigger sets and its noised version with the per-pixel perturbation bound of 16.
\old{It is evident that the perturbation is imperceptible.}\new{It is evident that the original content of the image is clearly perceptible.}
\marginpar{\margintext{Review-C-2}}
Table~\ref{tab:Amazon_result} in (Appendix~\ref{sec:confidence_score}) shows the original and our perturbed versions of some images from the same class of CelebA, and their corresponding confident scores from Amazon Rekognition API.

\begin{figure*}[ht]
    \centering
    \subfigure[CIFAR-10]{
    \includegraphics[width=.315\linewidth]{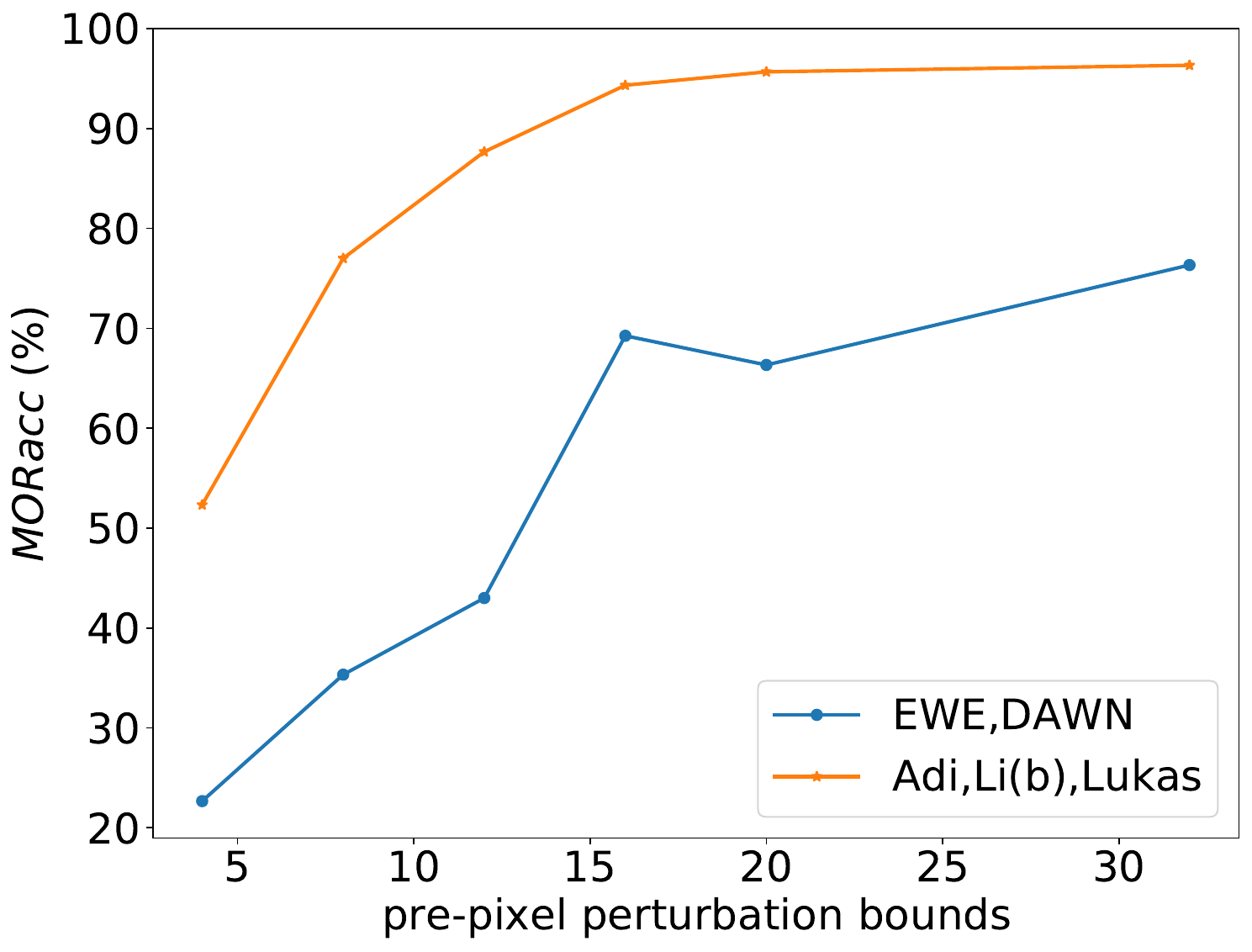}
    }
    \subfigure[ImageNet]{
    \includegraphics[width=.315\linewidth]{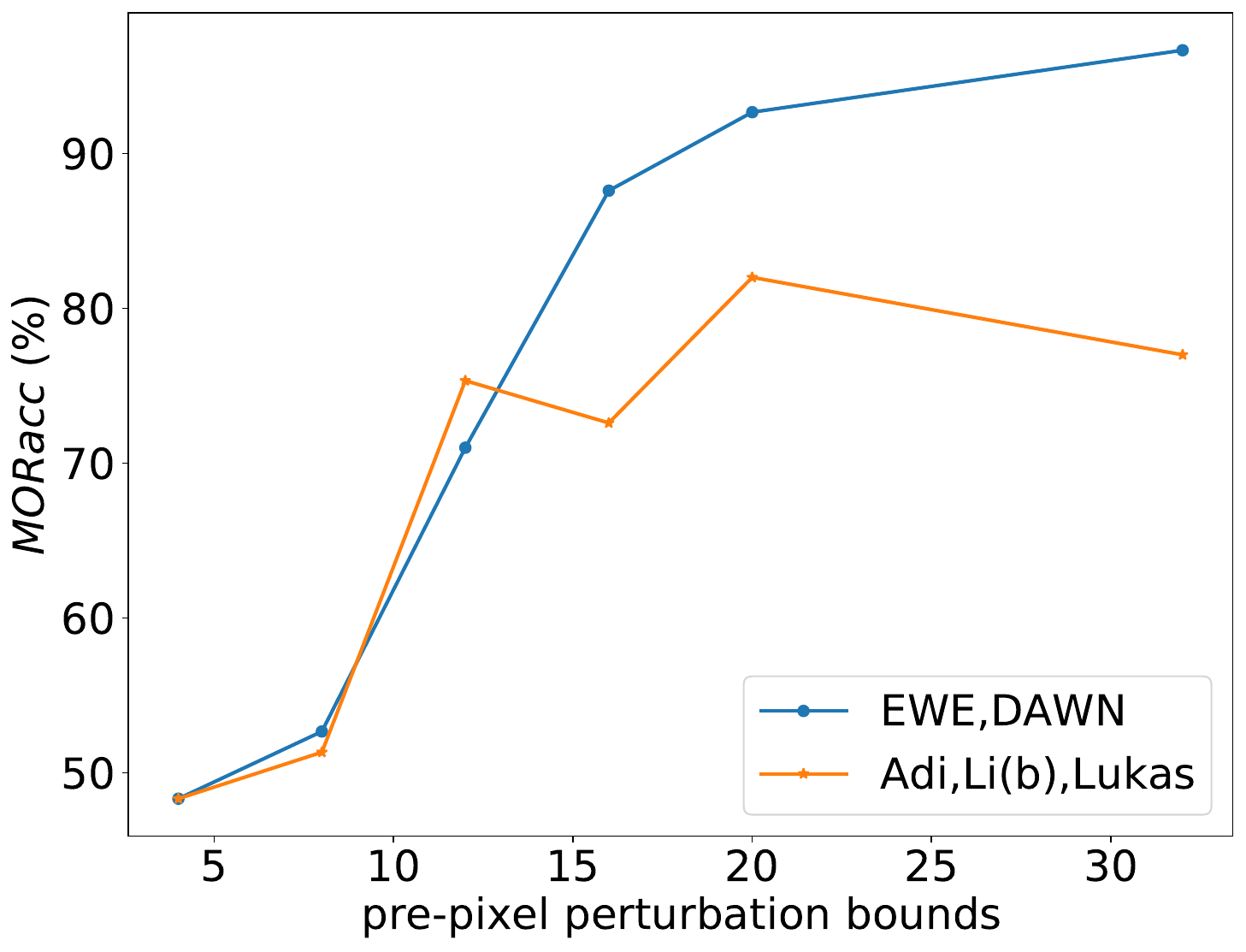}
    }
    \subfigure[CelebA]{
    \includegraphics[width=.315\linewidth]{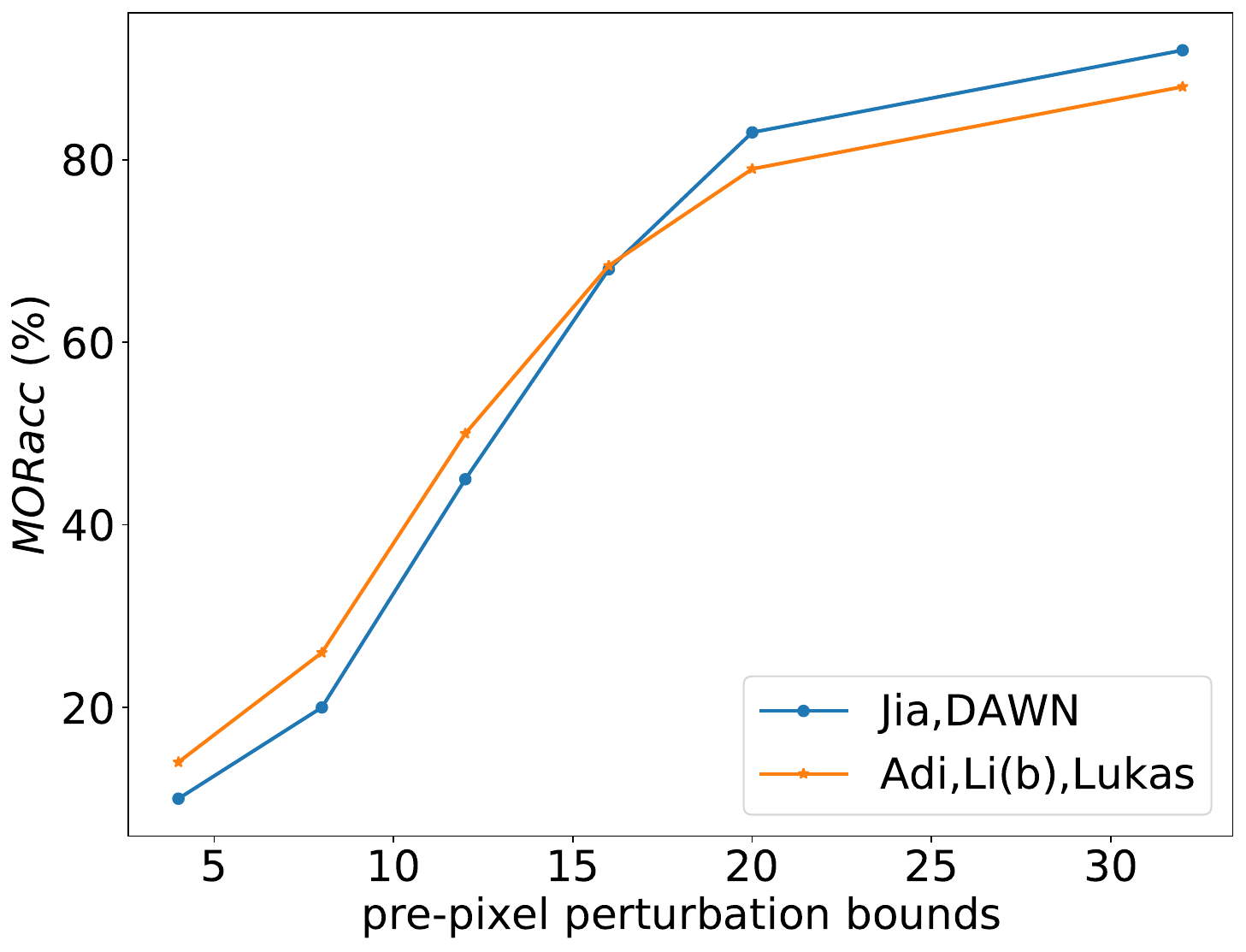}
    } 

    \caption{$\mathit{MORacc}$ with different per-pixel perturbation bounds.}
    \label{fig:eps}
\end{figure*}

We also measure the $\mathit{MORacc}$ with different per-pixel perturbation bounds. 
The results are in \fig~\ref{fig:eps}.
The $\mathit{MORacc}$ roughly increases with the perturbation bound and becomes stable when the $\mathit{MORacc}$ is high enough.
When generating untargeted adversarial examples on CIFAR-10, the untargeted label $y$ for all $F$s in Equation~\ref{equ:loss} tend to be the same one that near the true label.
As a result, the $\mathit{MORacc}$ of untargeted adversarial examples (Adi, Li (b) and Lukas) on CIFAR-10 are higher than the targeted ones (\old{Jia}\new{EWE} and DAWN).
This is not the case for ImageNet, as different $F$s have  different untargeted labels, leading to a relatively lower $\mathit{MORacc}$.
As there are only two labels in CelebA, the targeted and untargeted adversarial examples are essentially the same, 
hence, their $\mathit{MORacc}$ are simliar to each other. 

\revision{
Our attack can succeed even if $F_\accuser$ is of much lower quality than $F_\suspect$. 
We conducted additional experiments with CIFAR-10 and ImageNet to see what $F_\accuser$ accuracy is needed for a successful attack against an $F_\suspect$:
\begin{itemize}
    \item On CIFAR-10, $F_\suspect$ achieves 86.3\% accuracy; we can break the mix thresholds of all schemes with $F_\accuser$ reaching only 75.0\% accuracy; (with $F_\accuser$ reaching 80\% accuracy, we can break the extracted thresholds of all schemes except Li(b) and Lukas).
    \item On ImageNet, $F_\suspect$ achieves 85.9\%  accuracy; we can break the mix thresholds of all schemes with $F_\accuser$ reaching only 71.3\% accuracy; (with $F_\accuser$ reaching 76.3\% accuracy, we can break the extracted thresholds of all schemes except Li(b) and DAWN).
    \item While we do not know the precise accuracy of the Amazon Rekognition model, it undoubtedly surpasses that of our $F_\accuser$, given its status as a well-known commercial service.
\end{itemize}
}

\subsection{False claim efficiency}

Table~\ref{tab:falseclaimefficiency} displays the time usages in minutes for generating the adversarial trigger sets. Note that a malicious $\accuser$ needs to run this process only once; therefore the computational cost of our false claims is essentially negligible.

\begin{table}[ht]
\new{\caption{Attack runtime in minutes \revision{(the reported results does not include the time for training independent models;
training an independent model takes 18 min on CIFAR-10, 
11 min on ImageNet and 67.5 min on CelebA)}.\label{tab:falseclaimefficiency}}}
\marginnote{\margintext{Review-E-1}}
\small
\centering
\begin{spacing}{1.30}
\begin{tabular}{c|cccccc}
\toprule[1pt]
            &Adi   & \old{Jia}\new{EWE}   & Li (b) & DAWN  & Lukas & DI  \\
            \toprule[0.8pt]
 CIFAR-10   &13.6  & 14.1& 13.6 & 14.1 & 13.6 & 13.6  \\\hline
 ImageNet   &11.1  &19.1 & 11.1 &19.1 &11.1 & 11.1  \\\hline
 CelebA     &8.8  &10.7 &8.8  &10.7  &8.8  & 8.8\\
 \bottomrule
\end{tabular}
\end{spacing}

\end{table}

\section{Countermeasures}
\label{sec:countermeasures}

In this section, we present four kinds of countermeasures that have the potential to defend against our false claims. 
Three of them are generic and can be applied to all MOR schemes, while the fourth is specific to a particular scheme. 
We also examine the feasibility and potential pitfalls associated with each countermeasure.

\Paragraph{$\judge$-verified trigger sets.}
We remark that the root cause of our false claims is the ability of a malicious $\accuser$ to deviate from the trigger set generation procedure specified by a MOR scheme. 
Then, a straightforward countermeasure it to have $\judge$ verify if the trigger set was generated correctly.

At first glance, achieving this goal seems easy via {\em verifiable computation} (VC)~\cite{Pinocchio, libsnark}, which enables efficient verification of complex function execution through a VC-proof.
Namely, we could have $\accuser$ use VC to prove that the trigger set was indeed generated by following the MOR scheme.
Unfortunately, this is not applicable to watermarking, because its trigger set sampling cannot be captured by the VC-proof.
For example, in the claim generation of Adi (cf. Section~\ref{sec:Adi}), $\accuser$ could generate an adversarial $(\xx, \yy)$ in Step~1) and then generate the VC-proof from Step~2) onwards, allowing the false claim to succeed even with a valid VC-proof.
For fingerprinting, VC is applicable, since its trigger sets are generated based on $F_\accuser$.
To be secure, the VC-proof in fingerprinting must include the training stage of $F_\accuser$, otherwise, $\accuser$ could still generate an adversarial trigger set by manipulating the source model.
However, including the training stage in VC will be overwhelmingly expensive for the proof generation.
We further remark that VC is not applicable to DI due to the fact that $\Proof$-generation in DI relies not only on $F_\accuser$, but also on its training set, which can be manipulated in advance. 

A more effective way to verify the trigger set is to have $\judge$ train multiple independent models by itself and run them on the trigger set;
if $\judge$'s claim verification check deems all such known independent models as stolen, $\judge$ can reject $\Proof_\accuser$ as adversarial.
One caveat is that $\accuser$ could use black-box model extraction to extract these models and generate the trigger sets that can cause $\judge$ to deem them as independent;
$\accuser$ could also create black-box adversarial examples directly against these models.
While such attacks may be expensive and resource-intensive, it is crucial to take them into consideration. 
Therefore, $\judge$ has to 
rate-limit and/or raise the cost of dispute requests from $\accuser$s, which is reasonable since dispute rarely happens.

\Paragraph{$\judge$-generated trigger sets.}
Instead of verifying the trigger set, we could have $\judge$ \ul{generate} the trigger set by itself.
For model-independent watermarking, $\judge$ can simply choose a trigger set and return it to $\accuser$.
For model-dependent watermarking, 
$\accuser$ needs to submit $F_\accuser$ to $\judge$ first, who can then choose the trigger set based on $F_\accuser$.
In either case, $\accuser$ needs to interact with $\judge$ for each deployed model, even if no dispute happens.
For fingerprinting, it is enough for $\judge$ to get involved and generate the trigger set only when dispute happens.
It is worth mentioning that such a countermeasure is not applicable to DI or DAWN.
For DI, again, the training dataset can be manipulated so that $\Proof_\accuser$ can be adversarial even if it is generated by $\judge$.
For DAWN, the trigger set is generated dynamically based on the queries submitted by the clients.


\Paragraph{Defending against transferable adversarial examples.}
Notice that preventing false claims can be reduced to preventing transferable adversarial examples. 
If we have an effective way for detecting transferable adversarial examples~\cite{Delving, MIFGSM, DIFGSM}, we could have $\judge$ use it to determine if the trigger set is adversarial.
Similarly, if we have an effective way for adversarial training~\cite{shafahi2019adversarial, FGSM, bonawitz2019towards}, we could have $\suspect$ use it to make $F_\suspect$ more robust against transferable adversarial examples.
However, how to prevent transferable adversarial examples is still an open problem and it is orthogonal to our paper.
On the other hand, any advance in generating transferable adversarial examples will also enhance our false claims.

We evaluate the effectiveness of adversarial training against false claims (untargeted adversarial examples in particular) on CIFAR-10.
We use PGD~\cite{PGD} as the adversarial training algorithm and run it for 20 epochs with a perturbation bound $\epsilon=16$.
$MORacc$ decreases from $94.3\%$ to $32.5\%$.
However, the model accuracy also decreases from $86.3\%$ to $83.1\%$.
Therefore, adversarial training can be effective against false claims, but at the cost of a drop in utility.
\new{Therefore, adversarial training can be effective against false claims, but at the cost of a drop in utility. Furthermore, it takes 3-30 times longer to train a robust model with adversarial training than training a non-robust equivalent. This longer training time implies a higher cost in computational resources, which is particularly undesirable in the era of large models.}\marginpar{\margintext{Review-D-2}}
%
 %
However, note that $\accuser$ can use a large $\epsilon$ to invalidate adversarial training. 
In some settings,  $\judge$ can easily detect trigger set entries that are adversarial examples computed using a large $\epsilon$. How to use adversarial training as a broadly applicable, effective defense against false claim attacks remains an open problem. 
 
\Paragraph{Scheme-specific countermeasures.}
Recall that in DAWN, the trigger set consists of clients' queries;
if such queries are signed by clients, $\accuser$ can no longer perturb them.
However, $\accuser$ could collude with a client to submit perturbed queries.
To discourage such collusion, we could have $\judge$ punish the client who signed perturbed queries (instead of $\suspect$) when $F_{\suspect}$ is deemed stolen. This may not always be realistic.
Furthermore, signing each query separately is expensive. We would need to resort to having clients sign aggregate queries (e.g., in the form of a Merkle tree root) periodically.

\section{Discussion}
\label{sec:discuss}

\Paragraph{False claims for non-timestamped models.}
As we mentioned in Section~\ref{sec:mor_procedures}, most MOR schemes do not explicitly require the commitment to be timestamped. 
This is partially because timestamping is a strong assumption as it requires another trusted third party (or a blockchain) to run the service for timestamping.
To the best of our knowledge, none of the existing models is timestamped. 
That means we can falsely claim an existing model without considering timestamping. 

If timestamping is not enforced, our false claim becomes much easier, as $\accuser$ no longer has to create $\Proof_\accuser$ before $F_\suspect$ comes into existence.
That means $\accuser$ can directly query $F_\suspect$ to create adversarial examples, and such adversarial examples do not need to be transferable. 
As a result, some of the countermeasures presented in Section~\ref{sec:countermeasures} are no longer valid.
For example, we cannot rely on $\judge$ to train independent models to verify the trigger sets, because $\accuser$ can make the adversarial examples only work for  $F_\suspect$ (instead of transferring to other independent models) by leveraging the idea of Lukas~\cite{Lukas}.



\noindent\textbf{Deferred trigger set generation.}
Given that most model owners are unlikely to initiate MOR disputes, trigger set generation constitutes an unnecessary upfront cost, especially if we use the $\judge$-generated trigger set countermeasure (Section~\ref{sec:countermeasures}). 
Fortunately, in fingerprinting schemes, the trigger set is generated after $F_{\accuser}$ being fixed. 
Therefore, one can \emph{defer} trigger set generation until a MOR dispute arises. In this case, the certified timestamped commitment $cm_\accuser$ covers only $\accuser, F_{\accuser}$; $\judge$ will need to either generate or verify the trigger set in the event of a dispute.


\section{Related Work}
\label{sec:related}

\Paragraph{Parameter-Encoding.}
Recall that the main characteristic of our generalization (in Section~\ref{sec:mor_procedures}) is that $\Proof_\accuser$ includes a trigger set and its verification requires running model inference on the trigger set.
One kind of MOR schemes that do not follow our generalization is parameter-encoding-based watermarking, such as Uchida~\cite{Uchida}, DeepMarks~\cite{Deepmarks} and DeepSigns~\cite{Deepsigns}.  
Instead of embedding the watermark into the model functionality, they embed the watermark into the model parameters or the activations of its hidden layers;
and their claim verifications require accessing the model parameters.
For example, Uchida~\cite{Uchida} embeds a message into the weights of some target convolutional layer, by adding an embedding loss during training that regularizes the model and is minimized when the message can be extracted successfully.

Such schemes are known to be {\em not} robust~\cite{sok}:
one can easily remove the watermark via weight shifting, retraining, or transfer learning~\cite{wang2019attacks}.
Furthermore, when the suspect model is not white-box accessible, IP infringement cannot be detected.

\Paragraph{Passport-based watermarking.}
A special kind of parameter-encoding-based watermarking embeds a passport into the model to link the model to its owner's identity.
For example, Fan et al.~\cite{passport} introduce a passport layer to regulate the performance of the model: 
the model owner can use a secret passport to compute the affine factors of the passport layer. 
They claim that an attacker cannot add a substitute passport while maintain the model performance. 
However, Chen~\cite{passport_attack} et al. show that this is feasible, based on the observation that the model can achieve a high accuracy even with different affine factors of the passport layer. 
This attack can also be considered as a false claim, but they did not consider the situation where timestamped commitment is enforced.


We remark that, to link the model to its owner's identity, the owner can simply add its identity to the (timestamped) commitment, as we did in our generalization (in Section~\ref{sec:mor_procedures}).



\Paragraph{False positives in DI.}
Szyller et al.~\cite{szyller2022dirobustness} showed that DI suffers from false positives when an independent $F_{\suspect}$ happens to be trained from a dataset with the same distribution as $F_{\accuser}$'s training set.
This explains why DI has very high independent thresholds (cf. Table~\ref{tab:falseclaimeffectiveness}).
For example, for an honest trigger set, the normalized effective size of an independent model trained from CIFAR-10 can be as high as 90\%.

While Szyller et al.~\cite{szyller2022dirobustness} discovered the naturally occurring false positives, we further show that a {\em malicious} $\accuser$ can intentionally increase the normalized effective size to 100\%.




\Paragraph{Proof-of-learning.}
Jia et al.~\cite{PoL} proposed an alternative MOR scheme dubbed proof-of-learning (PoL), which allows $\accuser$ to claim ownership of its model by proving integrity of the training procedure.
Its $\Proof_\accuser$ includes a set of intermediate models recorded during training, together with the corresponding data points used to obtain each recorded model.
With such a $\Proof_\accuser$, $\judge$ can replicate the path all the way from the initial model to the final model to be fully confident that $\accuser$ has indeed performed the computation to obtain the final model.

Nevertheless, as a MOR scheme, PoL has three problems.
Firstly, PoL only helps if $F_\accuser$ and $F_\suspect$ are identical, i.e.,
it is not robust against the thief making minor changes to the stolen model.
Secondly, the authors of PoL only claimed that $\suspect$ cannot  generate a valid $\Proof_\accuser$ with a lower cost than that made by $\accuser$, i.e.,
it is not robust if $\suspect$ wishes to pay that much cost to generate $\Proof_\accuser$.
Lastly, Zhang et al.~\cite{zhang2022polspoof} have shown that, even with less cost than that made by $\accuser$, $\suspect$ can  still generate a valid $\Proof_\accuser$.

\section{Conclusion}
\label{sec:conc}
There has been a steady stream of works that have demonstrated that model theft is a real concern and MOR schemes are needed. The recent flurry of works that showed how large language models can be inexpensively bootstrapped off of prominent, expensive, models like GPT-4 have underscored this problem.

In this paper, we described a systematic shortcoming in all existing MOR schemes that permits false accusations to succeed, impacting the robustness, and hence the trustworthiness, of those schemes. Although we outlined several possible generic countermeasures, they can incur significant costs. Developing an efficient generic countermeasure, or identifying effective countermeasures specific to individual MOR schemes remain open problems.

\new{
Furthermore, we assume that the malicious accuser has access to a surrogate dataset that is in the same distribution as the suspect model’s training set, which may preclude false claims on valuable models trained on rare or private datasets. Evaluating the feasibility of false claims without this assumption is left for future work.}
\marginpar{\margintext{Review-A-2}}


\section*{Acknowledgments}
This work is supported in part by National Key Research and Development Program of China (2023YFB2704000), National Natural Science Foundation of China (U20A20222), Hangzhou Leading Innovation and Entrepreneurship Team (TD2020003), China Scholarship Council(CSC), Intel (in the context of the Private AI consortium), the Government of Ontario. Views expressed in the paper are those of the authors and do not necessarily reflect the position of the funders. We thank the anonymous reviewers, the program chair, and the shepherd for their constructive feedback and support. We also thank the authors of prior MOR schemes discussed here, who read our paper and provided valuable feedback.


\bibliographystyle{plain}

\bibliography{references}

\newpage

\appendix
\section{Experimental parameters}
\label{sec:parameters}

In this section, we present the parameters used in our experiments. Firstly we present the training settings of the DNN classifiers. Then we describe the parameters used in the FTAL attacks. Finally we describe the parameters for different MOR schemes.

\Paragraph{ImageNet} For Imagenet dataset, we select 10 classes,``goldfish'',``vulture'',``box\_turtle'',``clumber",``tabby'',
``cheetah'',``bassoon'',``cliff\_dwelling'',``wing'' and ``dough''.

\Paragraph{Training of DNN classifiers} We use batchsize 128 and SGD optimizer for all our trained classifiers.
For CIFAR-10, we train the models for 60 epochs with learning rate 0.1. 
For ImageNet, we train the models for 60 epochs with learning rate 0.01. 
For CelebA, we train the models for 10 epochs with  learning rate 0.1,

\Paragraph{Fine-Tune All Layers (FTAL)} We extract the trained classifier with their training set labeled by the classifier. We train the fine-tune models for 5 epoch with bathchsize 128. For CIFAR-10, we use the learning rate $lr\in [0.01,0.1]$. For ImageNet, the learning rate is $lr\in[0.001,0.01]$. And we use the learning rate $lr\in[0.01,0.1]$ for CelebA.

\Paragraph{Adi~\cite{Adi}} We use the Watermark-Robustness-ToolBox and concatenate the trigger to each batch of the training set. To achieve a $\mathit{MORacc}$ near 100\%, for CIFAR-10 and ImageNet, we concatenate 32 triggers to each batch. And for CelebA, we concatenate 2 triggers to each batch.

\Paragraph{\old{Jia} \new{EWE}~\cite{Jia}} We use the Watermark-Robustness-ToolBox. For CIFAR-10 and Imagenet, we use an SNNL weight 64 and 23, respectively, and a rate of 5, i.e. every fifth batch consist of watermark data. For CelebA, we use SNNL weight of 0.23 and a rate of 5.

\Paragraph{Li (b)~\cite{ASAC}} We set the batchsize as 100 for Li(b) for all three datasets.

\Paragraph{DAWN~\cite{DAWN}} We set the expected rate $r\approx 0.02$ at which a false label is returned.

\Paragraph{Lukas~\cite{Lukas}} We train 5 reference models and 5 surrogate models. And use IFGSM instead of PGD~\cite{PGD} (IFGSM has not random initialzation for the perturbation). We set the maximum value allowed for per-pixel perturbation as 16.

\Paragraph{DI~\cite{DI}}. We use their black box setting (Blind Walk) for CelebA since we do only has API access to Amazon Rekognition API. For CIFAR-10 and ImageNet, we use their white box setting (MinGD) to extract embeddings.

\section{Confidence scores}
\label{sec:confidence_score}

\begin{table}[ht]
\caption{Original images and adversarially perturbed images evaluated over Amazon Rekognition API. We report the confidence scores returned by Amazon Rekognition API.
\label{tab:Amazon_result}}
\centering
\begin{tabular}{|c|c|c|c|}
\hline
\begin{tabular}[c]{@{}c@{}}{\bf Original  image}\end{tabular}  & \begin{tabular}[c]{@{}c@{}}{\bf Female} \\ {\bf score} \end{tabular} & \begin{tabular}[c]{@{}c@{}}{\bf  Perturbed image}\end{tabular} & \begin{tabular}[c]{@{}c@{}}{\bf  Male} \\ {\bf score} \end{tabular} \\\hline

\raisebox{-.5\height}{\includegraphics[width=  0.19\linewidth]{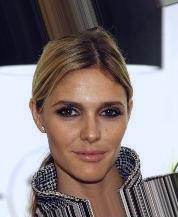}}                  &    99.8 \%                & 
\raisebox{-.5\height}{\includegraphics[width=0.19\linewidth]{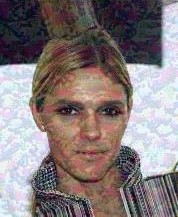}} &   99.9 \%                                                                                                                            \\\hline

\raisebox{-.5\height}{\includegraphics[width=0.19\linewidth]{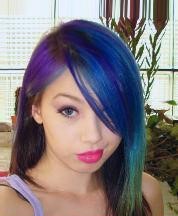}}  &    99.8 \%   & 
\raisebox{-.5\height}{\includegraphics[width=0.19\linewidth]{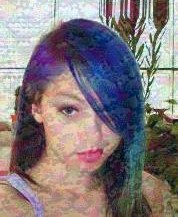}} &   88.3 \%     
\\ \hline  

\raisebox{-.5\height}{\includegraphics[width=0.19\linewidth]{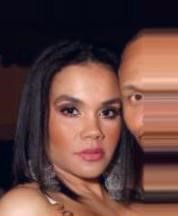}}  &    99.9 \%                             & 
\raisebox{-.5\height}{\includegraphics[width=0.19\linewidth]{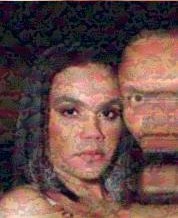}} &  
99.9 \%     
\\ \hline    
 
\raisebox{-.5\height}{\includegraphics[width=0.19\linewidth]{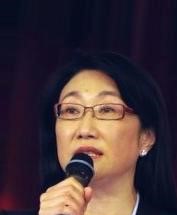}} &    99.9 \%                              & 
\raisebox{-.5\height}{\includegraphics[width=0.19\linewidth]{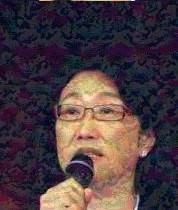}} &  
99.9 \%     
\\ \hline  
\raisebox{-.5\height}{\includegraphics[width=0.19\linewidth]{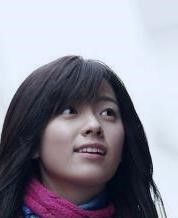}} &    99.9 \%                              & 
\raisebox{-.5\height}{\includegraphics[width=0.19\linewidth]{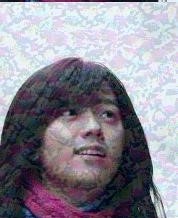}} &  
99.9 \%     
\\ \hline   
 
\end{tabular}
\setlength{\abovecaptionskip}{10pt}

\end{table}

\end{document}